\documentstyle[epsfig]{mn}
\newif\ifAMStwofonts
\newcommand{\be}{\begin{equation}}
\newcommand{\ee}{\end{equation}}
\newcommand{\bea}{\begin{eqnarray}}
\newcommand{\eea}{\end{eqnarray}}
\newcommand{\nn}{\nonumber}

\newcommand{\appgeq}{\stackrel{>}{\sim}}
\newcommand{\appleq}{\stackrel{<}{\sim}}

\newcommand{\Bphi}{B_\phi}
\newcommand{\Bz}{B_z}

\newcommand{\Bphis}{B_{\phi S}}
\newcommand{\Bzs}{B_{z S}}
\newcommand{\Ps}{P_S}      
\newcommand{\Rs}{R_S}

\newcommand{\sigsq}{\sigma^2}

\newcommand{\Pave}{\langle P \rangle}

\newcommand{\sigsqave}{\langle \sigma^2 \rangle}

\newcommand{\Gz}{\Gamma_z}
\newcommand{\Gphi}{\Gamma_\phi}

\newcommand{\Bo}{{\bf B_0}}
\newcommand{\Bi}{{\bf B_1}}
\newcommand{\vi}{{\bf v_1}}
\newcommand{\cross}{{\bf\times}}
\newcommand{\nhat}{{\hat n}} 
\newcommand{\mvir}{m_{vir}}
\newcommand{\G}{{\cal G}}

\title{Helical Fields and Filamentary Molecular Clouds II - Axisymmetric Stability and Fragmentation}
\author[J.D. Fiege and R.E. Pudritz]
       {Jason D. Fiege and Ralph E. Pudritz \\
	Dept. of Physics and Astronomy \\
	McMaster University \\
	1280 Main St. W.,
	Hamilton, Ontario \\
	L8S 4M1}
      
\pagerange{\pageref{firstpage}--\pageref{lastpage}}
\pubyear{1999}
       
\begin{document}

\maketitle
 
\label{firstpage}
 
\begin{abstract}
In Paper I (Fiege \& Pudritz, 1999), we constructed models of filamentary molecular clouds
that are truncated by a realistic external pressure and contain a rather general helical magnetic field.
We address the stability of our models to gravitational fragmentation and axisymmetric
MHD-driven instabilities.  By calculating the dominant modes of axisymmetric instability, we determine the
dominant length scales and growth rates for fragmentation.  
We find that the role of pressure truncation is to decrease the growth rate of
gravitational instabilities by decreasing the self-gravitating mass per unit length.
Purely poloidal and toroidal fields also help to stabilize filamentary clouds against fragmentation.   
The overall effect of helical fields is to stabilize gravity-driven modes, so that the growth rates
are significantly reduced below what is expected for unmagnetized clouds.
However, MHD ``sausage'' instabilities are triggered in models whose toroidal flux 
to mass ratio exceeds the poloidal flux to mass ratio
by more than a factor of $\sim 2$.
We find that observed filaments appear to lie in a physical regime where the growth rates of both
gravitational fragmentation and axisymmetric MHD-driven modes are at a minimum.
\end{abstract}
       
\section{Introduction}
\label{sec:intro}
It has long been known that filamentary molecular clouds often have nearly periodic density enhancements along
their lengths.  One has only to look at
the Schneider and Elmegreen (1979) catalogue of filaments to observe many examples of
so-called globular filaments, where quasi-periodic cores are apparent.
Such periodicity has also been noted along the central 
ridge of the $\int$-shaped filament of Orion A (Dutrey et al. 1991).  
The fragmentation of self-gravitating filaments has
accordingly received some attention.  
Given that overdense fragments may eventually evolve into star-forming
cores, it is important to have as complete a picture of fragmentation as possible.

Chandrasekhar and Fermi (1953) first demonstrated that uniform, incompressible filaments threaded by
a purely poloidal magnetic field are subject to gravitational instabilities, and that magnetic field helps to 
decrease the growth rate of the instability.
More recently, other authors (Nagasawa 1987, Nakamura, Hanawa, \& Nakano 1993, Tomisaka 1996, Gehman 1996, etc.)
have studied the stability of compressible, magnetized filaments.
We perform a stability analysis for the equilibrium models of filamentary 
molecular clouds that we described in Fiege \& Pudritz 1999 (hereafter FP1).
Our equilibria are threaded by helical magnetic fields and are truncated by the pressure of the ISM.
Most importantly, our models are constrained by observational data and are characterized by
approximately $r^{-2}$ density profiles, which are in good agreement with the observational data
(Alves et al. 1998, Lada, Alves, and Lada 1998).

We demonstrated in FP1 that most filaments are well below the critical mass per unit length required
for radial collapse to a spindle, and that any magnetized filamentary cloud that is initially in
a state of radial equilibrium is stable against purely radial perturbations.
Here, we consider more general modes of axisymmetric instability to find the growth rates and length scales
on which our models may break up into periodic fragments along the axis.
We show that our models are subject to two distinct types of instability.
The first is a gravity-driven mode of fragmentation, which occurs when the toroidal field component is relatively
weak compared to the poloidal.
The second type of instability arises when the toroidal magnetic field is relatively strong compared to the poloidal field,
and probably represents the axisymmetric ``sausage'' instability of plasma physics.
We find a stability criterion for these modes, which we discuss in Section \ref{sec:helix}.
We find that the gravitational and MHD modes
blend together at wavelengths
that are intermediate between the long wavelengths of purely gravitational
modes, and the very short wavelengths of purely MHD-driven instabilities.
These intermediate modes, in fact, represent a physical regime in which filamentary clouds fragment very slowly.

Once the process of fragmentation begins, it becomes non-linear on a 
timescale governed by the growth rate of the instability.
We stress that our analysis is linear and, therefore, can say nothing about the non-linear 
evolution of filamentary clouds.  However, the linear stability is important because it 
determines the dominant length scale for the separation of fragments, and the largest scale for 
substructure in filamentary clouds.  These fragments might eventually evolve into cores
In the present work, we limit our discussion to how the magnetic 
field configuration and external pressure affect spacing and mass scale of fragments.

We consider only axisymmetric modes here.  
The analysis of non-axisymmetric modes, most notably the
kink instability (cf. Jackson 1975), will be investigated in a separate paper.
One might expect fast-growing kink instabilities because of the toroidal character
of the outer magnetic field.  However, the centrally concentrated poloidal field is much stronger than the
peak toroidal field in most of our models; this magnetic ``backbone'' should largely stabilize
our models against the kink instability.

A brief plan of our paper is as follows.  In Section \ref{sec:formulation}, we 
discuss the equilibrium state of the molecular filament and atomic envelope,
as well the general method of our stability calculations.  
We derive the equations of motion and all boundary conditions in Sections \ref{sec:equations}
and \ref{sec:BCS}.  
We discuss the general stability properties of filaments in Section \ref{sec:disp}, and the test
problems that we used to verify our numerical code in Section \ref{sec:tests}.
We discuss the fragmentation of purely hydrodynamic filaments in Section \ref{sec:ost}, and 
filaments with purely poloidal or toroidal magnetic fields in Section \ref{sec:pure}.
In Section \ref{sec:helix}, we analyze the stability of the observationally constrained
models with helical fields discussed in Paper I.  Finally, in Section \ref{sec:discussion},
we discuss the significance of our results, and summarize our main findings.

\section{General Formulation of the Problem}
\label{sec:formulation}
The models described in FP1 involve
three parameters, two to describe the mass loading of the magnetic flux lines, and a third to 
specify the radial concentration of the filament.  
A Monte Carlo exploration of our parameter space led us to construct a class of 
magnetostatic models that are consistent with the observations.
We use the equations of linearized MHD to superimpose infinitesimal perturbations on these 
observationally allowed equilibria.
In general, we assume that molecular filaments are embedded in a less dense envelope of HI gas. 
In addition, we assume that the envelope is non-self-gravitating and 
is in equilibrium with the gravitational field of the filament.  Thus, the envelope is most dense at the 
interface between the molecular and atomic gas, and slowly becomes more rarified with radial distance.
There is, in fact, at least one known example of a filament that is embedded in molecular gas of lower density.
The $\int$-shaped filament of Orion A is a very dense filament that is presently fragmenting and forming 
stars.  Since its external medium is molecular, our treatment of the external medium as non-self-gravitating
strictly does not apply to this case.

We also assume that the poloidal field component ($\Bz$) remains constant in the HI envelope, while the toroidal  
field component ($\Bphi$) decays with radius as $r^{-1}$, so that the equilibrium state of the atomic
gas is current free.  We note that the magnetic field exerts no force in zeroth order in this configuration;
thus, equilibrium is determined by hydrostatic balance alone.  We model the HI envelope as a polytrope, which we discuss in Section
\ref{sec:envelope}.

Of the authors mentioned in section \ref{sec:intro}, only
Chandrasekhar and Fermi (1953) and Nagasawa (1987) considered filaments that are truncated at finite radius.
Chandrasekhar and Fermi considered the external medium to
be a vacuum, while Nagasawa considered an infinitely hot and non-conducting external medium
of zero density but finite pressure.
We treat the external medium as a perfectly conducting medium of finite density.
By using the most general possible boundary conditions between two perfectly conducting media, we 
self-consistently solve the equations of linearized MHD in both the molecular filament and the
surrounding HI envelope.  Our approach allows us to both study the effects of a finite density envelope on the 
predicted growth rates of the instability, and also to predict what gas motions might arise in the envelope
during fragmentation.

\subsection{The Polytropic HI Envelope}
\label{sec:envelope}
The analysis of Paper I provides the equilibrium state of the self-gravitating molecular filament.
However, we must also
find an equilibrium solution for the HI envelope.  
We assume that
the envelope is non-self-gravitating and current-free, with both field components continuous 
across the interface.  
Thus, the field in the envelope is given by
\bea
\Bz &=& \Bzs \nn\\
\Bphi &=& \Bphis \left(\frac{r}{\Rs}\right)^{-1},
\label{eq:HIfields}
\eea
where $\Rs$ is the radius of the molecular filament, as defined in Paper I, and
$\Bzs$ and $\Bphis$ are the field components at the surface.

It is required that the the total pressure of the gas and magnetic
field balance at the interface.  Thus the boundary condition on the total pressure is given by
\be
\left[P+\frac{\Bz^2+\Bphi^2}{8\pi}\right]=0,
\ee
where we use square brackets to denote the jump in a quantity across the interface.
Since we have assumed that the field components are continuous in our model, 
the pressure must be continuous also.

The magnetic field exerts no force on the gas in our current-free configuration.
Therefore, we may derive the equilibrium structure of the envelope
by taking the gas to be in
purely hydrostatic balance with the gravitational field of the molecular filament;
\be
0=\frac{dP_e}{dr}+\rho_e\frac{d\Phi}{dr},
\label{eq:eq1}
\ee
where subscript {\em e} refers to the HI envelope, and all quantities have their usual meanings.  
We assume that the gas
is polytropic with polytropic index $\gamma_e$;
\be
\frac{P_e}{P_S}=\left(\frac{\rho_e}{\rho_{e,S}}\right)^{\gamma_e}.
\label{eq:polytrope}
\ee
The gravitational potential outside of a cylindrical filament, with mass per unit length $m$, is given by
\be
\Phi=2Gm\ln{\frac{r}{\Rs}},
\label{eq:phi}
\ee
where we choose the zero point of $\Phi$ to be at the surface of the filament (radius $\Rs$).
With the help of \ref{eq:polytrope} and \ref{eq:phi}, it is easy to solve for the density structure
of the HI envelope:   
\be
\rho_e = \frac{P_S}{\sigsq_{e,S}}\left(1-2\frac{\gamma_e-1}{\sigsq_e\gamma_e}Gm\ln{\frac{r}{R_S}}\right)^
        \frac{1}{\gamma_e-1}.
\label{eq:rho}
\ee
The pressure distribution may then be obtained from equation \ref{eq:polytrope}.
We find that the pressure falls very slowly with radius, especially for $\gamma_e<1$.  Generally
we will take $0.25\le\gamma_e\le 0.5$ to simulate an atomic envelope that is nearly 
isobaric out to very large radius.

\section{The Equations of Linearized MHD}
\label{sec:equations}
In this Section, we set up the eigensystem of equations governing the motion of a perfectly conducting, self-gravitating,
magnetized gas.  
We assume that all perturbed variables can be expressed as Fourier modes.  For example, we write the perturbed density $\rho_1$ 
in the form
\be
\rho_1(r,z,t)=\rho_1(r) e^{i(\omega t+m\phi+k_z)},
\ee
where $m=0$ for the axisymmetric modes considered in this paper.  
We note that all quantities are written in dimensionless form using the dimensional scalings 
introduced in Paper I.

We shall customarily reserve the subscript ``0'' for equilibrium quantities and ``1'' quantities in first
order perturbation.  In Appendix \ref{app:MHD}, we show that the linearized equations of MHD reduce to the 
following set of coupled differential equations.  The momentum equation combined with induction and continuity equations becomes
\bea
-\omega^2\rho_0\vi &=& \gamma\nabla\cdot(\rho_0\vi)      \nn\\
&+& \frac{1}{4\pi}\left\{(\nabla\cross\Bo)\cross\left[\nabla\cross\left(\rho_0\vi\cross\frac{\Bo}{\rho_0}\right)\right]  \right. \nn\\
&+& \left. \left[\nabla\cross\nabla\cross\left(\rho_0\vi\cross\frac{\Bo}{\rho_0}\right)\right]\cross\Bo\right\} \nn\\
&-& \rho_0\nabla\cdot(i\omega\Phi_1)+\nabla\Phi_0\nabla\cdot(\rho_0\vi),
\label{eq:eig1}
\eea
and Poisson's equation becomes
\be
0=\nabla^2\varphi_1+\G\nabla\cdot(\rho_0\vi).
\label{eq:eig2b}
\ee
where $\varphi_1=i\omega\Phi_1$.
The adiabatic index $\gamma$ is allowed to be discontinuous at the interface between the molecular and atomic gas; in general we define
\be
\gamma=\left\{  	\begin{array}{cc}
			1, & r\le\Rs \\
			\gamma_e, & r>\Rs 
			\end{array}
	\right\}.
\ee
We have also included a parameter $\G$ in equation \ref{eq:eig2b} that allows us to ``turn off'' self-gravity in
the HI envelope, as discussed in section \ref{sec:formulation}:
\be
\G=\left\{  	\begin{array}{cc}
			1, & r\le\Rs \\
			0, & r>\Rs 
			\end{array}
	\right\}.
\ee.

Equations \ref{eq:eig1} and \ref{eq:eig2b} represent the eigensystem, written entirely in terms
of the perturbed momentum density $\rho_0 v_1$ and $\varphi_1$.
Appendix \ref{app:MHD} shows that equations \ref{eq:eig1} and \ref{eq:eig2b} can be finite differenced
and written in the form of a standard matrix eigenvalue problem:
\be
-\omega^2{\bf\Psi}={\hat L}\Psi,
\label{eq:EIG}
\ee
where $\hat L$ is a $4\times 4$ block matrix operator, and
$\Psi$ is an ``eigenvector'' made up of the components of the momentum density and the 
modified gravitational potential $\varphi$:
\be
{\bf \Psi}=\left[       \begin{array}{c}
                        \rho_0 v_{r1} \\
                        \rho_0 v_{\phi 1} \\
                        \rho_0 v_{z1} \\
                        \varphi_1
                        \end{array}
           \right].
\label{eq:Psi}
\ee
It has been shown by Nakamura (1991) that the equations of self-gravitating, compressible, ideal MHD are self-adjoint; thus
the eigenvalue $-\omega^2$ of equation \ref{eq:EIG} is purely real, and $\omega$ may be purely real or purely
imaginary. When $\omega$ is real, the perturbation is stable, since it 
just oscillates about the equilibrium state.  On the other hand, when $\omega$ is imaginary,
the perturbation grows exponentially, and the system is unstable.  It is these unstable modes that we are primarily concerned with, since they
lead to fragmentation.

\section{Boundary Conditions}
\label{sec:BCS}
Equation \ref{eq:EIG} does not yet fully specify the eigensystem since boundary conditions have yet to be imposed.
There are three radial boundaries in our problem; the first two are obviously the boundaries that occur at $r=0$ and
$r\rightarrow\infty$.  The third is the internal boundary that that separates the molecular filament and the 
HI envelope.  As we discussed in Section \ref{sec:equations}, we cannot directly difference across the interface
because the density is discontinuous on that surface.  Instead, we specify the appropriate boundary conditions, thus 
linking the perturbation of the HI envelope to that of the filament.

\subsection{The Inner Boundary: $r=0$}
\label{sec:BC1}
The boundary condition at the radial centre of the filament ($r=0$) is trivial for axisymmetric modes ($m=0$).
Mass conservation demands that $\rho_0 v_r\rightarrow 0$, and the perturbed gravitational field $d\Phi_1/dr\rightarrow 0$
since the internal mass per unit length vanishes as $r\rightarrow 0$.

\subsection{The Outer Boundary: $r\rightarrow\infty$}
\label{sec:BC2}

The outer boundary conditions are that all components of the perturbation vanish at infinity; 
thus, all components of the momentum density $\rho_0\vi$, and the perturbed potential $\varphi_1$,
must vanish.

It would be difficult to specify these boundary conditions on a uniform grid.
Therefore, we employ a non-uniform radial grid spacing defined by the transformation
\be
r=S \tan{\xi}
\ee
where $S$ is a constant scale factor; thus radial infinity is mapped to $\xi=\pi/2$.  In practice, we vary
$\xi$ over the range $\delta_1\le\xi\le\pi/2-\delta_2$ for small $\delta_1$ and $\delta_2$,
which define the inner and outer boundaries.  Thus, our numerical grid extends fron nearly zero radius to essentially 
infinite radius.  Our transformation has the benefit of excellent dynamic range; by choosing $S$ appropriately, we can arrange to
have most of the grid points (approximately $2/3$) inside of the molecular filament, where good resolution is most important,
with progressively fewer points in the HI envelope as $r\rightarrow\infty$.  
In practice, we find that the eigensystem is quite insensitive to the choice of $\delta_2$; there is no
detectable change in the eigenvalue $-\omega^2$ for $\delta_2$ correspanding to outer radii greater than $\sim 100 ~r_0$,
where $r_0$ is the core radius defined in Paper I for filamentary clouds.

\subsection{The Molecular Filament/HI Envelope Internal Boundary}
\label{sec:BC3}
The most complex boundary in the problem occurs at the interface between the molecular and atomic gas.  This surface is
a contact discontinuity which moves freely, but through which no material may pass.  No material may cross the 
boundary since this would involve a phase transition between atomic and molecular gas, which requires
much longer than the dynamical timescales relevant to our problem.
The boundary conditions at the interface are written as follows:
\bea
~[\Phi] &=& 0 \label{eq:bc1} \\
~[\nabla\Phi\cdot\nhat] &=& 0 \label{eq:bc2}\\
~[{\bf B}\cdot\nhat] &=& 0 \label{eq:bc3} \\
~\left[-\left(P+\frac{ {\bf B}\cdot{\bf B} }{8\pi}\right)\nhat+\frac{\bf B}{4\pi}\left(\nhat\cdot{\bf B}\right)\right] &=& 0 \label{eq:bc4},
\eea
where square brackets denote the jump in a quantity across the boundary.  We note that these jump conditions
apply in a Lagrangian frame that is co-moving with the deformed surface. 
Equations \ref{eq:bc1} and \ref{eq:bc2} demand that the gravitational potential and its first derivative
(the gravitational field) must be continuous aross the interface.  Equation \ref{eq:bc3} is the usual condition on the
normal magnetic field component from electromagnetic theory; it is easily derived from the divergence-free condition of the 
magnetic field.  The final condition, equation \ref{eq:bc4} states that the normal component of the total stress is
continuous across the boundary.

Appendix \ref{app:BC} derives the explicit forms of these equations in terms of the components of $\Psi$ (equation \ref{eq:Psi}),
and Appendix \ref{app:matrix} shows how the boundary conditions can be included in the matrix eigensystem 
given by equation \ref{eq:EIG}.
Equation \ref{eq:EIGFINAL} gives
the final form of our eigensystem, which we solve, in Section \ref{sec:disp}, for various equilibrium configurations.
In general, equation \ref{eq:EIGFINAL} has $4\times(N-1)$ 
eigenvalues, where $N$ is the size of each block matrix,
but most of these represent stable MHD waves.  We generally solve only for the dominant mode, 
since fragmentation will be dominated by the 
fastest growing axisymmetric instability (with largest positive $-\omega^2$).
For this reason, our problem is well suited for iterative methods.  We solve equation \ref{eq:EIGFINAL}
using the well-known method of shifted inverse iteration (cf. Nakamura 1991).  

\section{Dispersion Relations and Eigenmodes}
\label{sec:disp}
For any equilibrium state (which may be prepared using the formulation of Paper I),
we may determine a dispersion relation for the dominant mode by solving equation \ref{eq:EIGFINAL} for $-\omega^2$
as we vary the wave number $k_z$.  The wave number specifies the wavelength $\lambda$ over which the instability operates;
\be
\lambda=\frac{2\pi}{k_z}.
\label{eq:lambda}
\ee

We shall be primarily concerned with the following characteristics of our dispersion curves.
1) $k_{z,max}$ is the wavenumber of maximum instability, which may determine the scale for the
separation of fragments by equation \ref{eq:lambda}.
2) $-\omega^2_{max}$ is the maximum squared growth rate for unstable modes, and determines the timescale on
which the instability operated.
3) $k_{z,crit}$ is the maximum wave number for which a mode is unstable.  Therefore, it sets the minimum
length scale of the instability.
We shall often write $k_{z,max}$, and $k_{z,crit}$, and $-\omega^2_{max}$ in dimensionless form, defined by
\bea
\tilde{ k}_{z,max} &=& r_0 k_{z,max} \nn\\
\tilde{ k}_{z,crit} &=& r_0 k_{z,crit} \nn\\
-\tilde{\omega}^2_{max} &=& -\omega^2_{max} (4\pi G \rho_c)^{-1}),
\label{eq:dimensionless}
\eea
where tildes are reserved for dimensionless quantities for the remainder of this paper.

\section{Tests of the Numerical Code}
\label{sec:tests}
We have tested our code by reproducing some of the dispersion curves given by Nagasawa (1987) and Nakamura (1993).
These test problems include the following: 1) the untruncated Ostriker solution (Nakamura), 2) untruncated
filaments with purely poloidal, purely toroidal, and helical fields (Nakamura), and 3) pressure truncated
filaments with constant poloidal fields (Nagasawa).

Nagasawa assumes that the external medium is non-conducting and infinitely hot, with constant pressure and 
zero density.  Our results are indistinguishable from his when we use his boundary conditions.  When we include 
the effects of perfect MHD (infinite conductivity) and finite density in the external medium, our results converge
towards his in the limit of high velocity dispersion $\sigma_e \appgeq 10-100$.  At lower, and more
realistic, velocity dispersions ($\sigma_e \approx 5$), we find that the instability is slightly decreased, although the
general character of his curves is preserved.  These effects are further discussed in Section \ref{sec:ost}.

\section{Stability of the Pressure Truncated Ostriker Solution}
\label{sec:ost}
In this Section, we systematically examine the effects of the density and pressure of the external medium
on the stability of hydrostatic filaments.
The equilibrium state is the isothermal Ostriker solution discussed in Paper I;
\be 
\rho=\frac{\rho_c}{\left[1+(r/r_0)^2\right]^2},
\ee
which may be truncated at any desired radius by external pressure (see Paper I).  

\begin{figure}
\epsfig{file=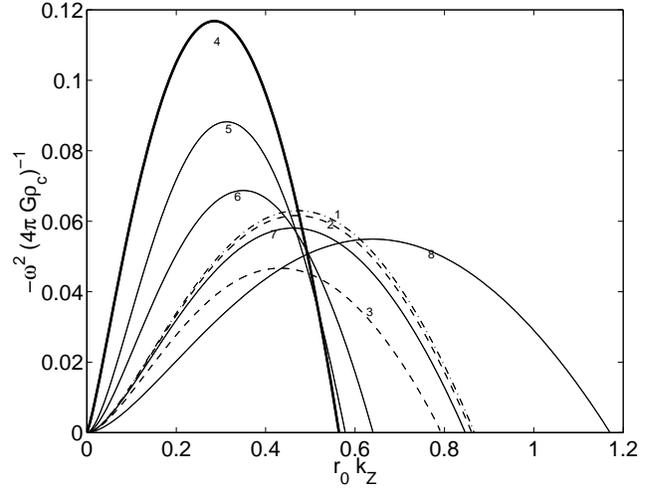,width=\linewidth}
\caption{We have computed dispersion curves for truncated, unmagnetized, isothermal filaments.  The external medium
has finite density and a velocity dispersion of $\sigma_e=5\sigma$ for the solid curves.  The curve shown as dot-dashed 
has an infinitely hot, zero density, external medium.  The dashed curves have been computed using external media with 
velocity dispersions $\sigma_e=10\sigma$ for the upper curve, and $\sigma_e=2.5\sigma$ for the lower curve. 
Table \ref{tab:fig1} summarizes the results shown here.}
\label{fig:Ost}
\end{figure}

Two independent effects are shown in Figure \ref{fig:Ost}; curves 1 to 3 show the effect of varying the
density of the external medium, while the solid curves (curves 4 to 8) shows the effect of varying the external
pressure, or equivalently the mass per unit length.  

We first consider the effect of varying the density of the HI envelope.
Equation \ref{eq:rho} gives the density, as a function of radius, for our polytropic envelope.  We note that the density 
is controlled, mainly, by the velocity dispersion $\sigma_{e,S}$ just outside of the molecular filament.
The dashed curves (1 to 3) in Figure \ref{fig:Ost} have been computed using using the same equilibrium solution for the
molecular filament (see Table \ref{tab:fig1}), but different values for $\sigma_{e,S}$.  For all three curves, we have assumed that
$\Ps/\Pave=0.801$ and $m/\mvir=0.199$.
We find that envelopes with lower velocity dispersions and higher densities have a slightly stabilizing effect.
Although we have taken the external medium to be non-self-gravitating, it responds dynamically and self-consistently to the
gravitational field and motions of the molecular filament.
As the filament fragments, the deformation of the surface and the changes
in the gravitational potential induce motions in the surrounding gas.  These motions slightly resist the fragmentation of the filament,
because of the finite inertial mass of the envelope.
We have also computed one solution, shown as the curve numbered 1,
where we have assumed that the external medium is infinitely hot, and of infinitely low density, which is
consistent with Nagasawa's (1987) treatment of the boundary conditions (see Section \ref{sec:BCS}).  It is apparent that
dispersion relations computed with fully self-consistent boundary conditions converge to this curve in the
limit of high velocity dispersion.  In practice, we find essentially no difference when $\sigma_e/\sigma_c \appgeq 10$.

We now consider the effect of the external pressure, or equivalently, the effect of the mass per unit length.
The virial equation from Paper I
demonstrates that pressure truncation is equivalent to a reduction of the mass per unit length for hydrostatic filaments;
\be
\frac{m}{\mvir}=1-\frac{\Ps}{\Pave},
\ee
where $\mvir=2\sigsqave/G$ is the virial mass per unit length.
Thus, $\Ps/\Pave$ and $m/\mvir$ cannot be varied independently for unmagnetized filaments.
The solid lines in Figure \ref{fig:Ost} (curves 4 to 8) show dispersion relations for the truncated Ostriker solution,
where we have assumed that the velocity dispersion of the HI envelope just outside of the filament
is five times that of the molecular gas (see Table \ref{tab:fig1}). 
We find that more severely pressure truncated filaments, 
with higher $\Ps/\Pave$ and 
lower $m/\mvir$, are less unstable than more extended filaments.  Thus, pressure truncation suppresses the gravitational instability of the 
filament.  This result is easy to understand. Decreasing the mass per unit length decreases the gravitational force that drives the instability.  
Therefore, pressure dominated filaments, with external pressure comparable to the central pressure, are more stable than those 
that are dominated by self-gravity (eg. untruncated filaments).

\begin{figure}

\epsfig{file=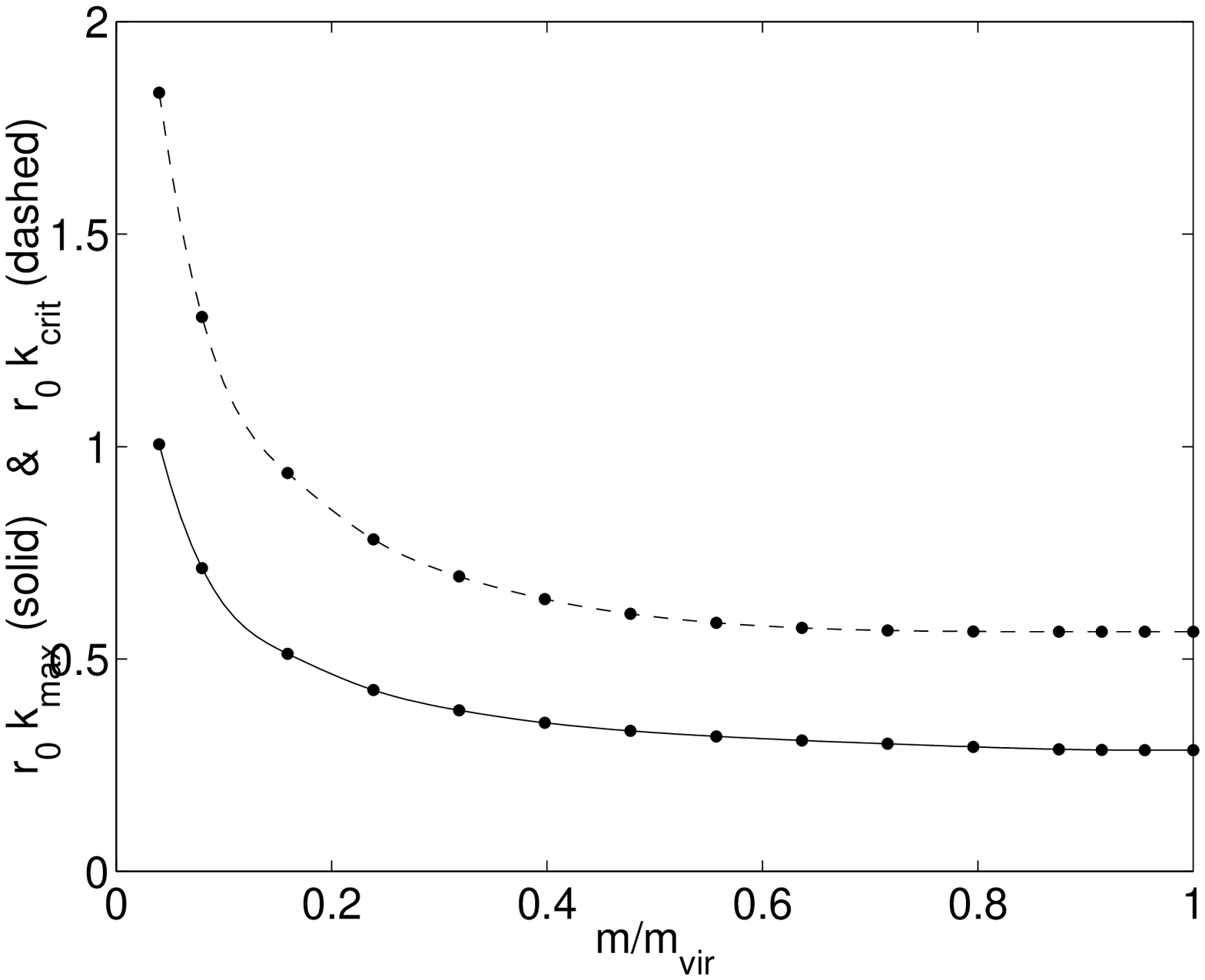,width=\linewidth}

\epsfig{file=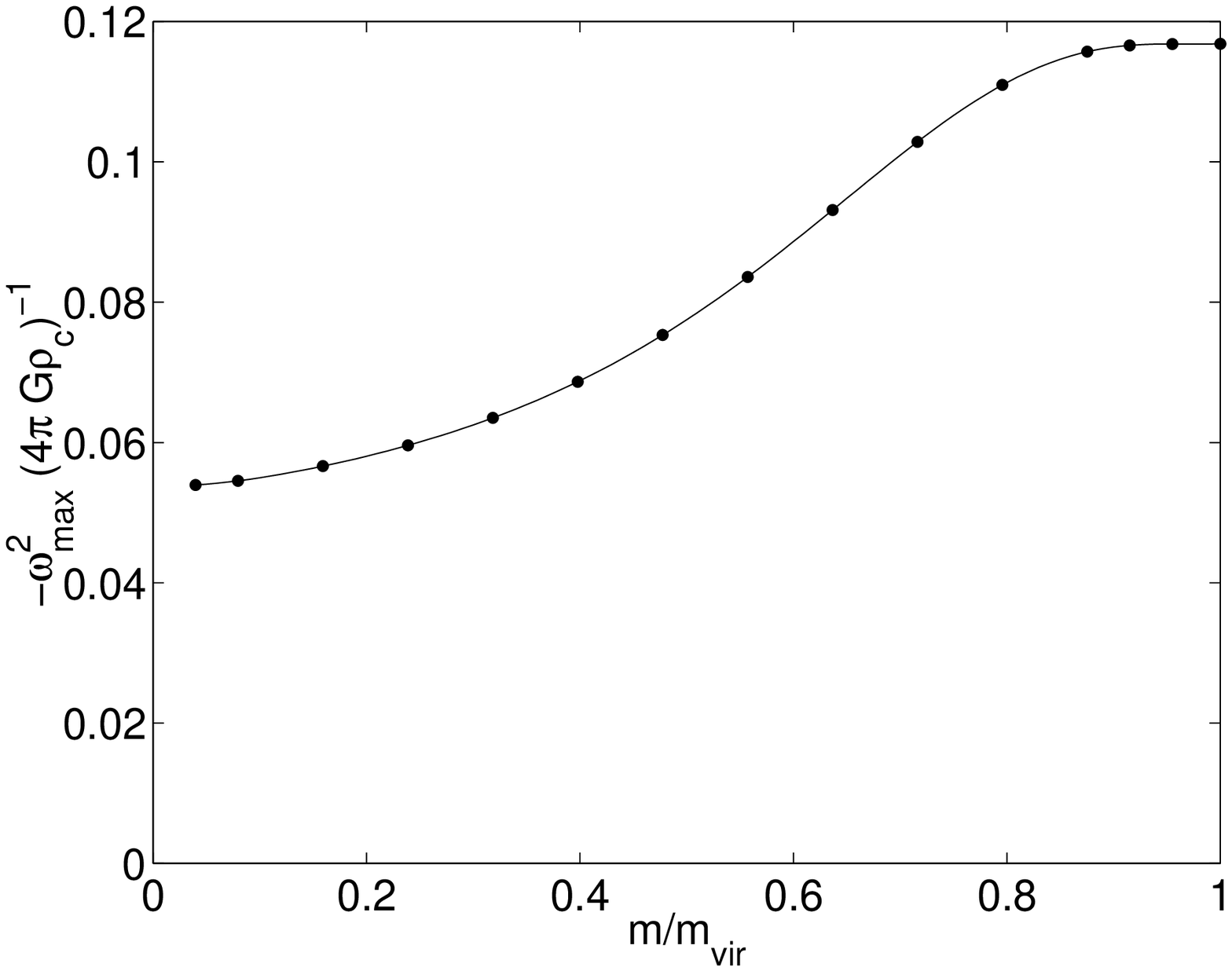,width=\linewidth}
\caption{We plot the wave numbers $k_{z,max}$ and $k_{z,crit}$, as well as the squared growth rate $-\omega^2$ of the most 
unstable mode, as a function of the $m/\mvir$.}
\label{fig:cont1}
\end{figure}

\begin{table}
\begin{minipage}{\linewidth}
\begin{tabular}{l|c|c|c|c|c|c}
\hline
  & $m/m_{vir}$ & $P_S/\Pave$ & C  & $\tilde{k}_{z,max}$ & $-\tilde{\omega}_{max}^2  $ & $\tilde{k}_{z,crit}$ \\ 
\hline
1 &  0.199    & 0.801    & 0.472    & 0.149    & 0.063    & 0.866  \footnote{Infinitely hot external medium.}\\
2 &  0.199    & 0.801    & 0.47     & 0.149    & 0.0616   & 0.861  \footnote{``Hot external medium'': $\sigma_e=10\sigma$} \\
3 &  0.199    & 0.801    & 0.432    & 0.149    & 0.0467   & 0.791  \footnote{``Cold external medium'': $\sigma_e=2.5\sigma$} \\
4 &  1        & 0    	 & $\infty$ & 0.285    & 0.117    & 0.564  \\
5 &  0.597    & 0.403    & 0.537    & 0.313    & 0.0882   & 0.578  \\
6 &  0.398    & 0.602    & 0.362    & 0.35     & 0.0687   & 0.64   \\
7 &  0.199    & 0.801    & 0.149    & 0.462    & 0.0581   & 0.847  \\
8 &  0.0995   & 0.901    & -0.027   & 0.639    & 0.0549   & 1.17   \\
\hline
\end{tabular}
\end{minipage}
\caption{We give the the wave numbers and growth rates for the fastest growing modes of truncated, unmagnetized filaments
(See Figure \ref{fig:Ost}).  We also give the critical growth rate, beyond which the filament is 
stable against axisymmetric perturbations.  The tildes indicate dimensionless variables, as defined by equations
\ref{eq:dimensionless}.  The external medium has finite density and a velocity dispersion of $\sigma_e=5\sigma$, 
except where otherwise indicated in the footnotes.}
\label{tab:fig1}
\end{table}   

\begin{figure}
\epsfig{file=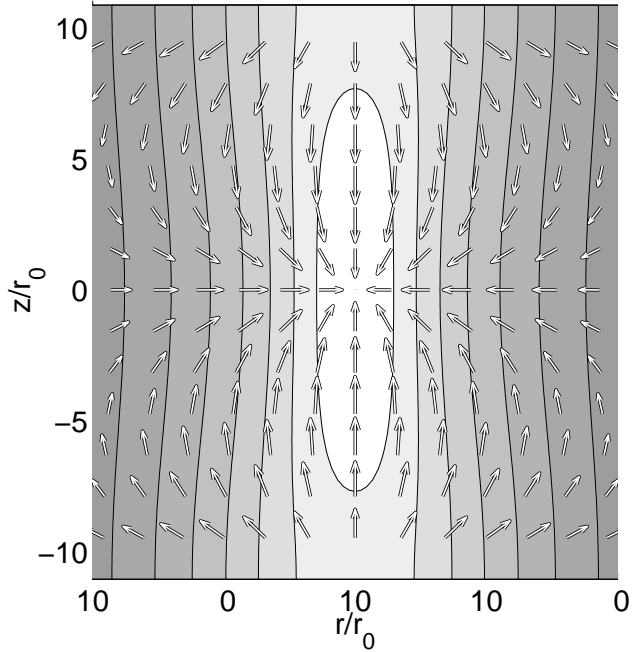,width=\linewidth}
\caption{We show the eigenmode corresponding to the most unstable mode
of the untruncated Ostriker solution ($r_0 k_{z,max}=0.285$, $-\omega^2 (4\pi G\rho_c)^{-1}=0.117$).  The contours show the density, 
while the arrows show the velocity field.  We note that the segment shown is just one wavelength in the periodically fragmenting
filament.  The arrows are scaled logarithmically with the velocity, as discussed in the text.}  
\label{fig:Osteig1}
\end{figure}

\begin{figure}
\epsfig{file=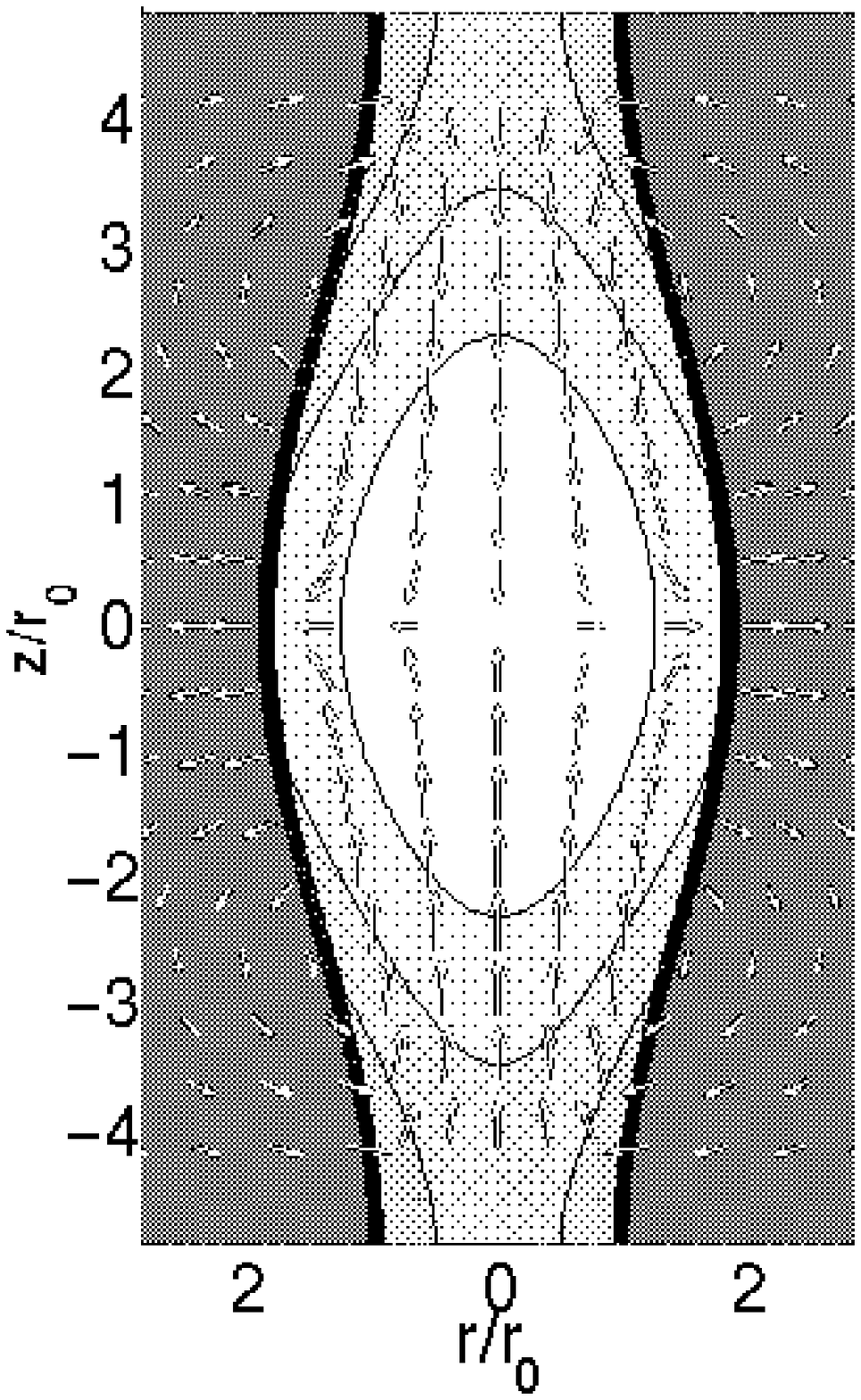,width=\linewidth}
\caption{We show the eigenmode corresponding to the most unstable mode
of a truncated Ostriker model with mass per unit length
$m/m_0=5$ ($m/\mvir=0.199$). For this mode, $r_0 k_{z,max}=0.462$ and $-\omega^2 (4\pi G\rho_c)^{-1}=0.449$.}
\label{fig:Osteig2}
\end{figure}

Figure \ref{fig:cont1} shows how the mass per unit length affects the
wave number $k_{z,max}$ and growth rate $-\omega_{max}^2$ of the most unstable mode, 
as well as the critical wave number $k_{z,crit}$.  We find that
the growth rate is significantly suppressed when $m/\mvir\appleq 0.8$.
The wave numbers remain essentially constant until 
$m/\mvir\approx 0.3$.  For smaller masses per unit length, the wave number increases, corresponding to fragmentation on smaller
length scales.

Two examples of eigenmodes are shown in figures \ref{fig:Osteig1} and \ref{fig:Osteig2} for unmagnetized filaments; 
the first shows the fragmentation of the untruncated Ostriker solution, while the second 
is significantly truncated ($m/\mvir=0.199$). These figures show a single wavelength in the periodic
fragmentation of a filament. 
We note that the size of the perturbation is 
greatly exaggerated in these figures, where we typically add a perturbation of $\approx 33\%$ (in peak density) to the unperturbed
solution.  

The arrows in figures \ref{fig:Osteig1} and \ref{fig:Osteig2} show the velocity field,
but the scaling is logarithmic due to the large range of velocities that must be shown.  Specifically, $arrow~length \propto
log_{10}(1+\alpha v/\sigma_c)$, where we choose $\alpha\approx 50$.  Both cases have purely poloidal velocity fields.
The untruncated case shows an asymmetric infall onto the fragment, with almost purely radial infall near the fragments,
and some radially outward motions between them.
On the other hand, the truncated model shows infall along the axis, but the motions are actually radially outward in the vicinity
of the fragment.  This is due simply to the outward bulging of the fragment, which can occur because there is no infalling material
approaching from the radial direction.

There are interesting motions in the gas outside of the truncated
filament shown in Figure \ref{fig:Osteig2}.  We observe inward motions,
that are driven by external pressure crushing the less dense parts of the filament between fragments.  At the same time, the bulging of the 
filament near the fragments drives a radially outward flow in the 
surrounding gas.  We emphasize that no mass is actually exchanged between the molecular
filament and the surrounding atomic gas.
Gas flows towards the fragments along the filament axis, which then bulge outwards 
slightly.  This induces circulation motions in the surrounding envelope in which
gas flows away from the fragments and towards the increasingly evacuated inter-fragment regions.

\section{Stability of Truncated Filaments with Purely Poloidal and Purely Toroidal Fields}
\label{sec:pure}
Following Paper I, we study the effects of poloidal and toroidal fields
separately before considering the more general case of helical
fields in Section \ref{sec:helix}.  We note that none of the equilibrium models
used in this Section fall within the range of observationally allowed
models from Paper I.
These models are useful, however, in that they provide insight into the roles
played by the field components in more general helical field models.

\subsection{Poloidal Field Results}
\label{sec:pol}
\begin{figure}
\epsfig{file=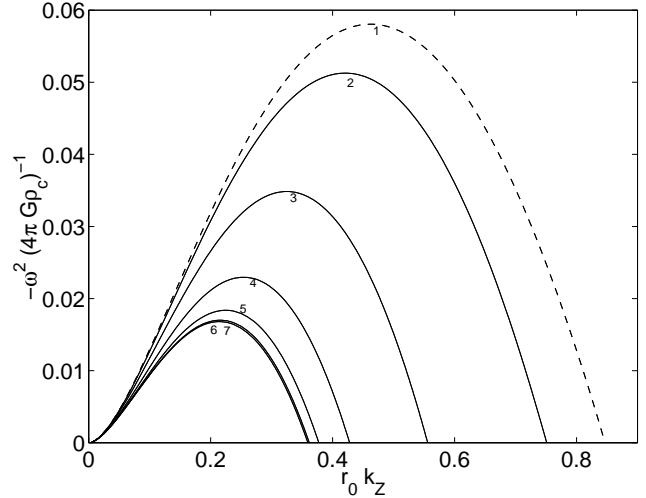,width=\linewidth}
\caption{We have computed dispersion curves for truncated filaments with poloidal field only.  In all cases,
the mass per unit length is $5 m_0$ ($m/\mvir=0.199$), which falls within the range of observationally allowed 
masses per unit length found in Paper I.  We have also assumed an external medium which is perfectly conducting
and of finite density, with $\sigma_e=5 \sigma$.
The labeling of the dispersion curves corresponds to the numbering in Table \ref{tab:fig2}.}
\label{fig:pol}
\end{figure}

\begin{figure}
\epsfig{file=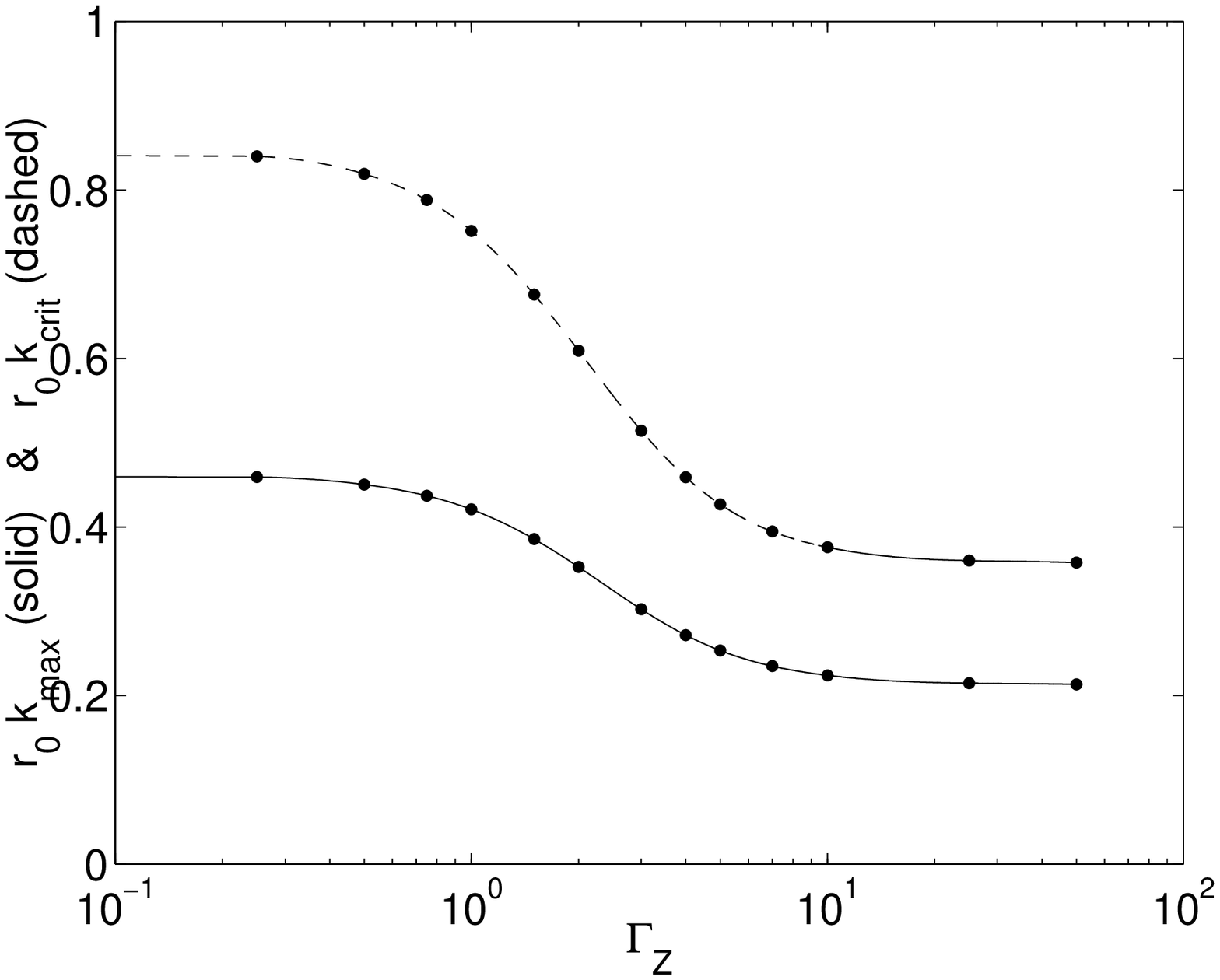,width=\linewidth}

\epsfig{file=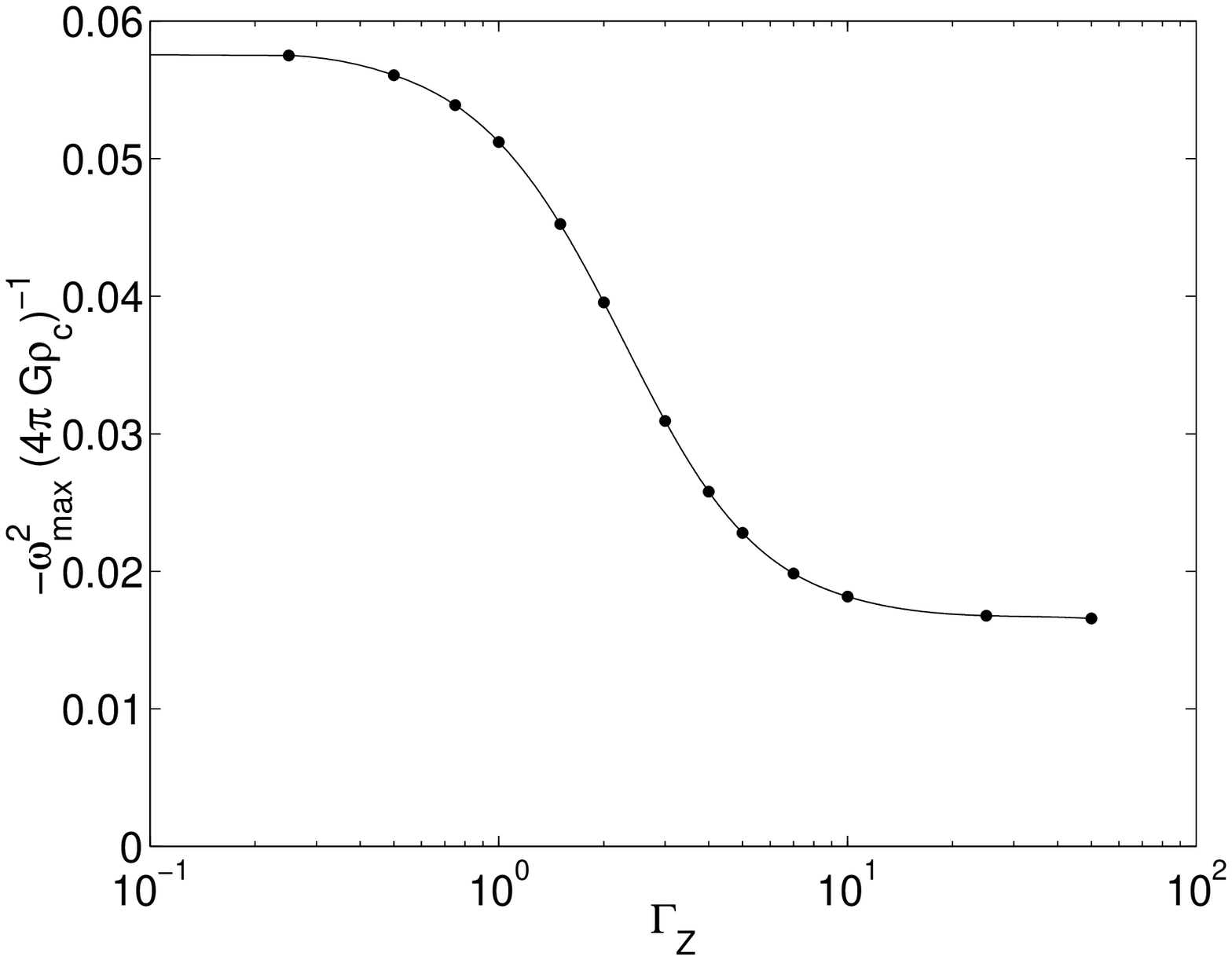,width=\linewidth}
\caption{We plot the wave numbers $k_{z,max}$ and $k_{z,crit}$, as well as the squared growth rate $-\omega^2$ of the most
unstable mode, as a function of the $\Gz$.}
\label{fig:contpol}
\end{figure}

Figure \ref{fig:pol} shows a sequence of dispersion curves, in which we
have varied the poloidal flux to mass ratio $\Gz$ while holding the
mass per unit length constant at $m=5~m_0$ ($m/\mvir=0.199$).
We observe that the poloidal field has a stabilizing effect on the
filament, but that the stabilization saturates for $\Gz\appgeq 10$.
The critical and most unstable wave numbers are decreased by the poloidal field,
but this effect also saturates when  $\Gz\appgeq 10$.
These results agree qualitatively with those obtained by Nagasawa (1987) for 
pressure truncated isothermal filaments threaded by purely poloidal fields, although the 
degree of stabilization is greater in our model.
Chandrasekhar and Fermi (1953) also find that poloidal fields stabilize incompressible 
filaments, but they do not find any saturation of the stabilizing effect. 
This is especially apparent in Figure \ref{fig:contpol}, where we have
plotted $k_{z,max}$, $k_{z,crit}$, and $-\omega_{max}^2$ as a funtion of $\Gz$.

\begin{figure}
\begin{minipage}{.49\linewidth}
\epsfig{file=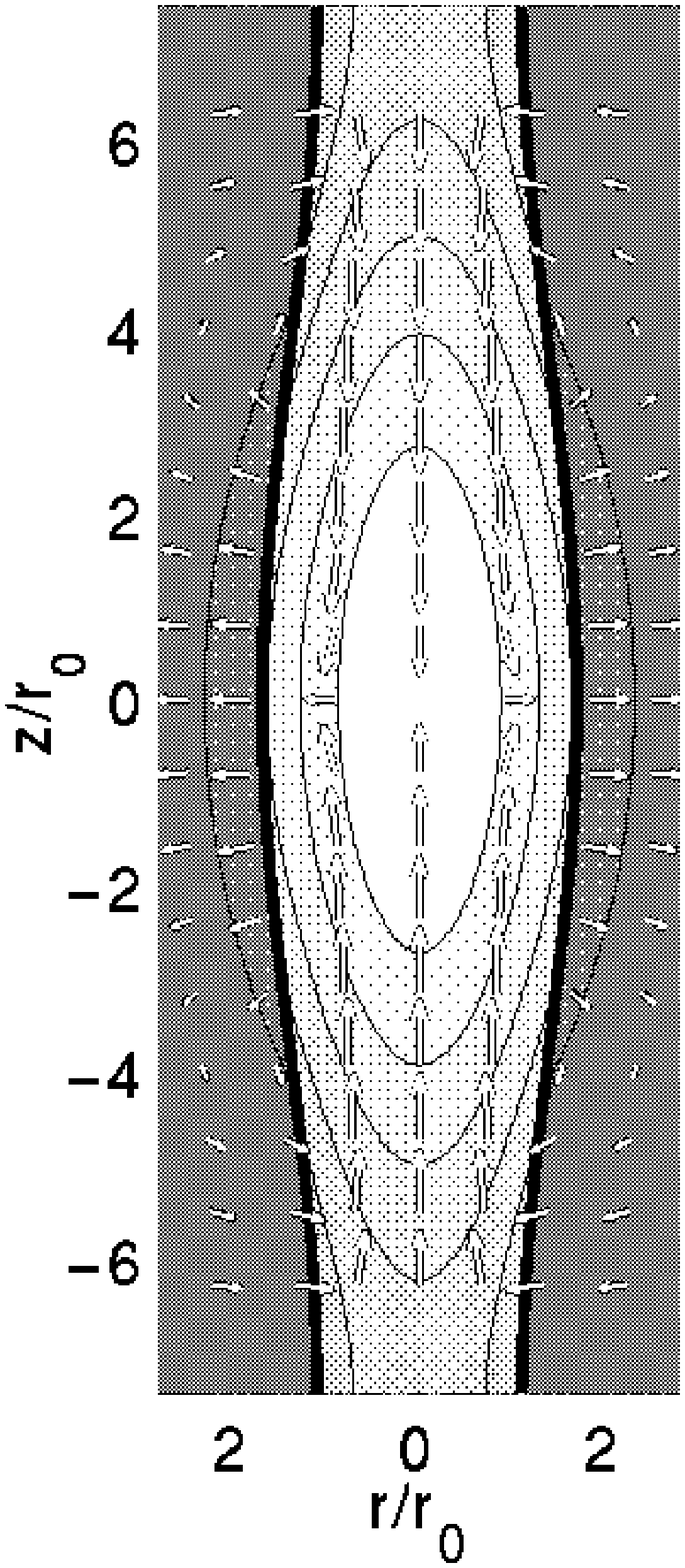,width=\linewidth}
\end{minipage}
\begin{minipage}{.49\linewidth}
\epsfig{file=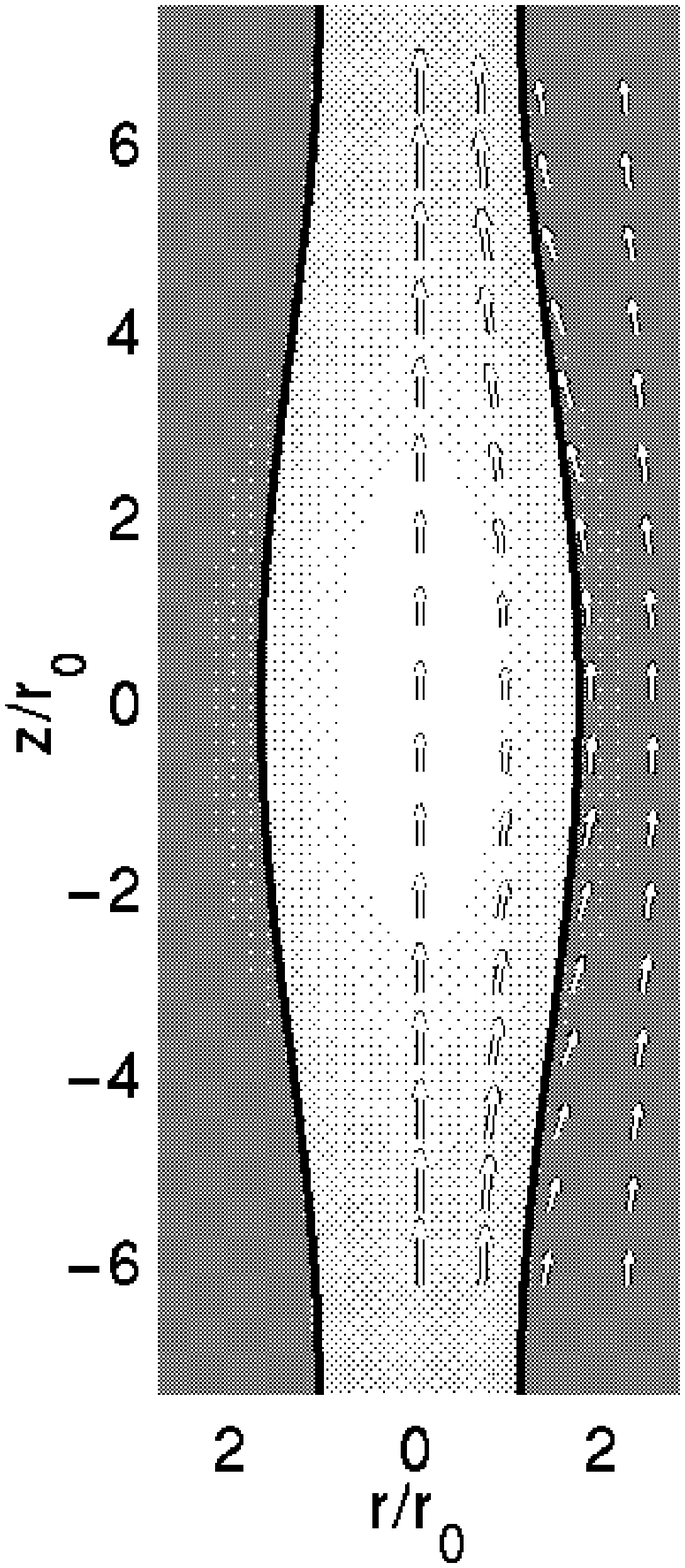,width=\linewidth}
\end{minipage}
\caption{We show the eigenmode corresponding to the most unstable mode
of a truncated magnetized model with a weak poloidal field ($\Gz=1$), and mass per unit length
$m/m_0=5$ ($m/\mvir=0.199$). For this mode, $r_0 k_{z,max}=0.421$ and $-\omega^2 (4\pi G\rho_c)^{-1}=0.0512$.  The figures represent a) (left panel) density
contours with superimposed poloidal velocity vectors.  We note that there is no toroidal velocity when the field is purely poloidal.
and b) (right panel) The magnetic field.  We always represent the magnetic field as
a split-frame figure, with poloidal field vectors shown on the right, and contours to represent the toroidal field on the left.  Since no toroidal
field is generated in this case, the left side of the figure is left blank.} 
\label{fig:polweakeig}
\end{figure}

\begin{figure}
\begin{minipage}{.49\linewidth}
\epsfig{file=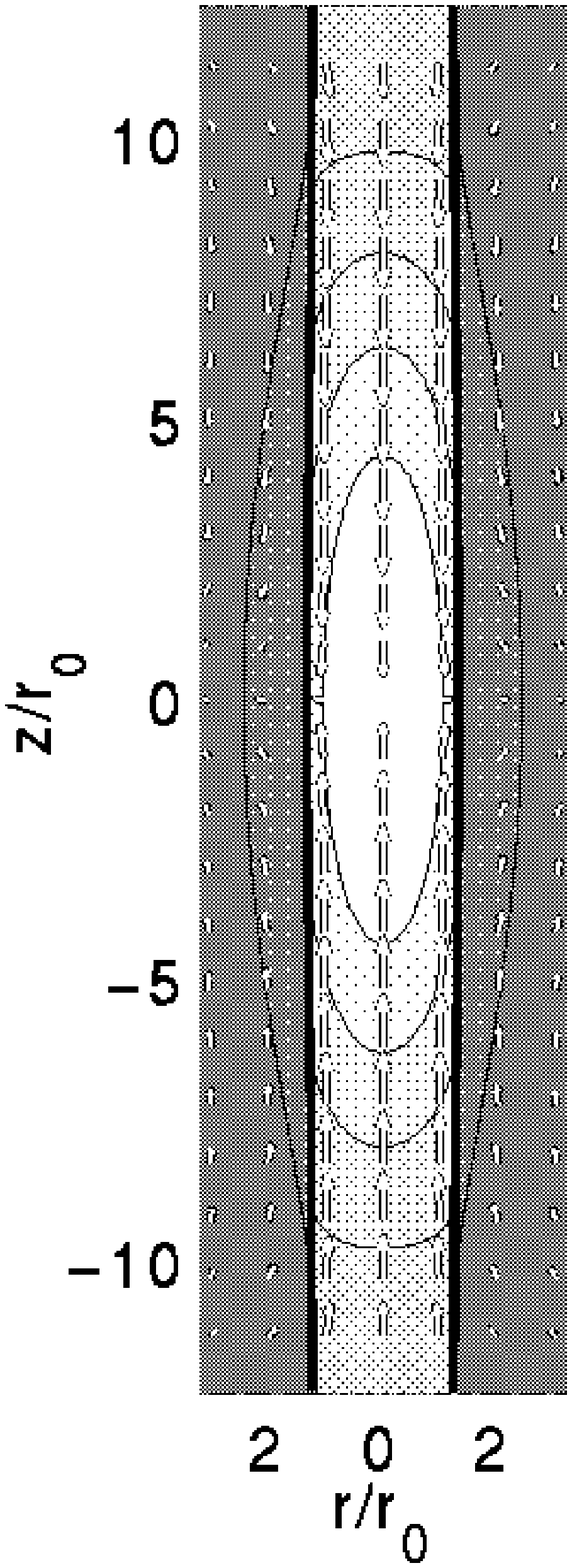,width=\linewidth}
\end{minipage}
\begin{minipage}{.49\linewidth}
\epsfig{file=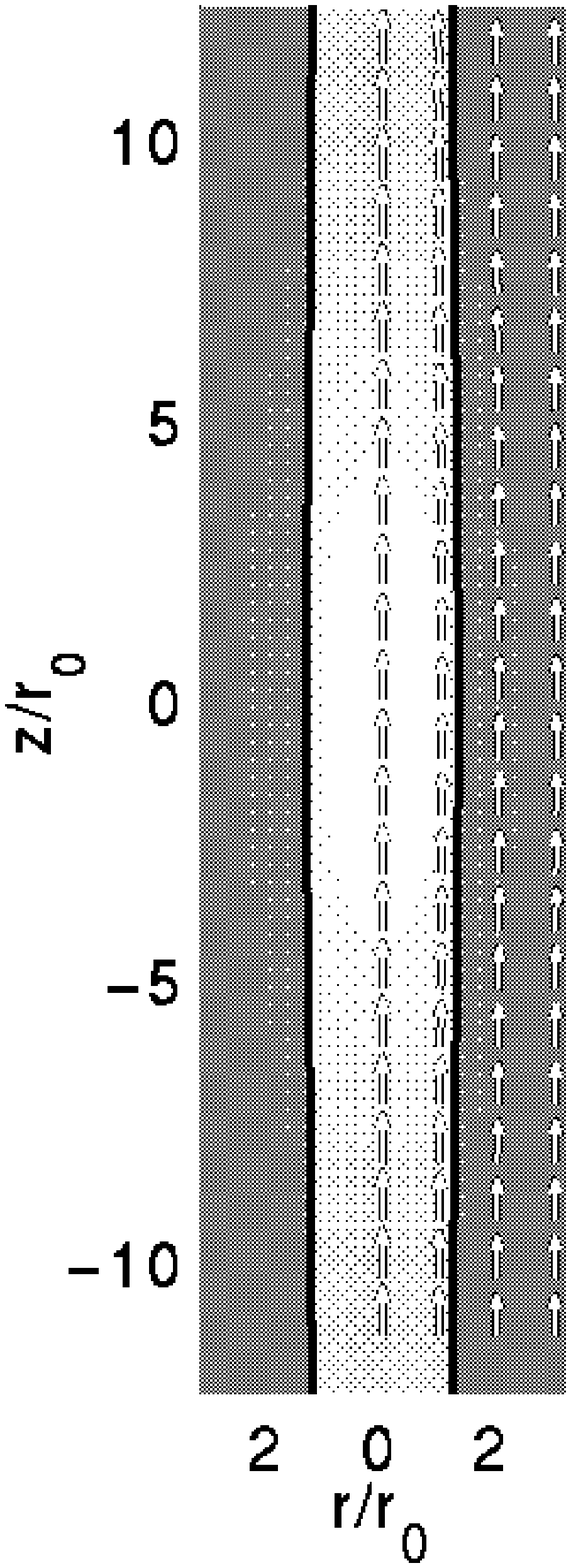,width=\linewidth}
\end{minipage}
\caption{We show the eigenmode corresponding to the most unstable mode
of a truncated magnetized model with a relatively strong poloidal field ($\Gz=5$), and mass per unit length
$m/m_0=5$ ($m/\mvir=0.199$). For this mode, $r_0 k_{z,max}=0.225$ and $-\omega^2 (4\pi G\rho_c)^{-1}=0.0228$.  
The figures represent a) (left panel) density
contours with superimposed poloidal velocity vectors.  We note that there is no toroidal velocity when the field is purely poloidal.
and b) (right panel) The magnetic field.  We always represent the magnetic field as
a split-frame figure, with poloidal field vectors shown on the right, and contours to represent the toroidal field on the left.  Since no toroidal
field is generated in this case, the left side of the figure is left blank.}
\label{fig:polstrongeig}
\end{figure}

Figures \ref{fig:polweakeig} and \ref{fig:polstrongeig} show two eigenmodes corresponding, respectively, to 
a weak ($\Gz=1$) and relatively strong ($\Gz=5$) poloidal field.  
Naturally, as the poloidal flux to mass ratio
increases, and the field strengthens, the gas becomes more tightly constrained to move 
only along the field lines.  When $\Gz \appgeq 10$, the motions are almost entirely poloidal; 
further increasing $\Gz$ has no significant effect on the motions, and, hence, no significant effect
on the dispersion curves.

\begin{table}
\begin{tabular}{l|c|c|c|c|c|c}
\hline
     & $\Gz$ & C & $P_S/\Pave$ & $\tilde{k}_{z,max}$ & $-\tilde{\omega}_{max}^2 $ & $\tilde{k}_{z,crit}$ \\
\hline
1    & 0    & 0.149   & 0.801    & 0.462   & 0.0581    & 0.846 \\
2    & 1    & 0.146   & 0.813    & 0.421   & 0.0513    & 0.751 \\
3    & 2.5  & 0.133   & 0.859    & 0.325   & 0.0348    & 0.556 \\
4    & 5    & 0.116   & 0.929    & 0.254   & 0.023     & 0.428 \\
5    & 10   & 0.106   & 0.977    & 0.225   & 0.0184    & 0.377 \\
6    & 25   & 0.102   & 0.996    & 0.216   & 0.017     & 0.362 \\
7    & 50   & 0.101   & 0.999    & 0.214   & 0.0168    & 0.359 \\
\hline
\end{tabular}
\caption{We give the the wave numbers and growth rates for the fastest growing modes of 
truncated filaments with poloidal field only (See Figure \ref{fig:pol}).  
We also give the critical growth rate, beyond which the filament is stable against axisymmetric
perturbations.  In all cases,
the mass per unit length is $5 m_0$ ($m/\mvir=0.199$), which falls within the range of observationally allowed 
masses per unit length found in Paper I.  We have also assumed an external medium which is perfectly conducting
and of finite density, with $\sigma_e=5 \sigma$.}
\label{tab:fig2}
\end{table}

\subsection{Toroidal Field Results}
\label{sec:tor}
\begin{figure}
\epsfig{file=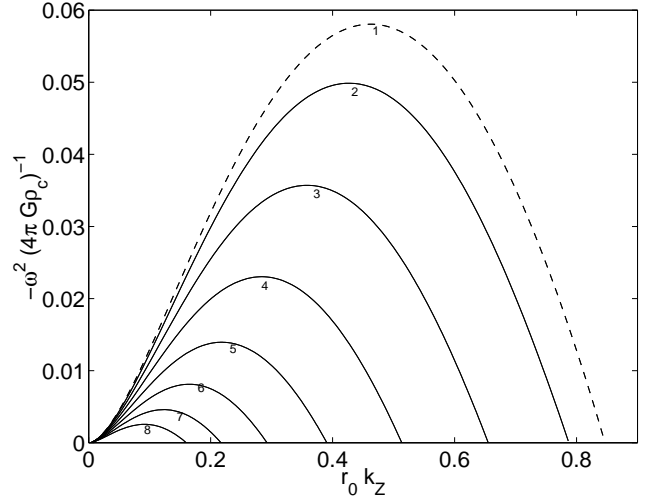,width=\linewidth}
\caption{We have computed dispersion curves for truncated filaments with toroidal field only.  In all cases,
the mass per unit length is $5 m_0$ ($m/\mvir=0.199$), which falls within the range of observationally allowed
masses per unit length found in Paper I.  We have also assumed an external medium which is perfectly conducting
and of finite density, with $\sigma_e=5 \sigma$.
The labeling of the dispersion curves corresponds to the numbering in Table \ref{tab:fig3}.}
\label{fig:tor}
\end{figure}

\begin{figure}
\epsfig{file=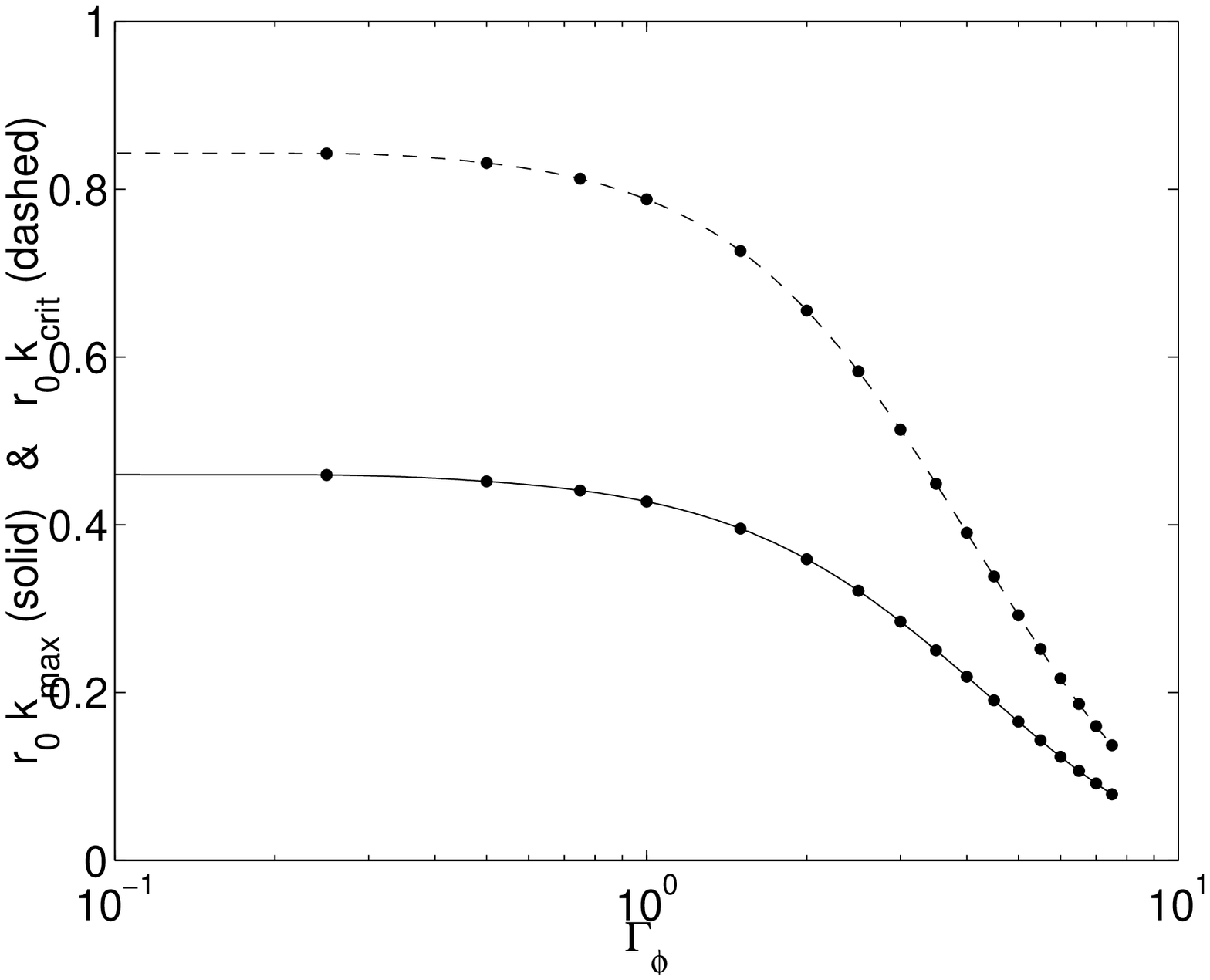,width=\linewidth}  

\epsfig{file=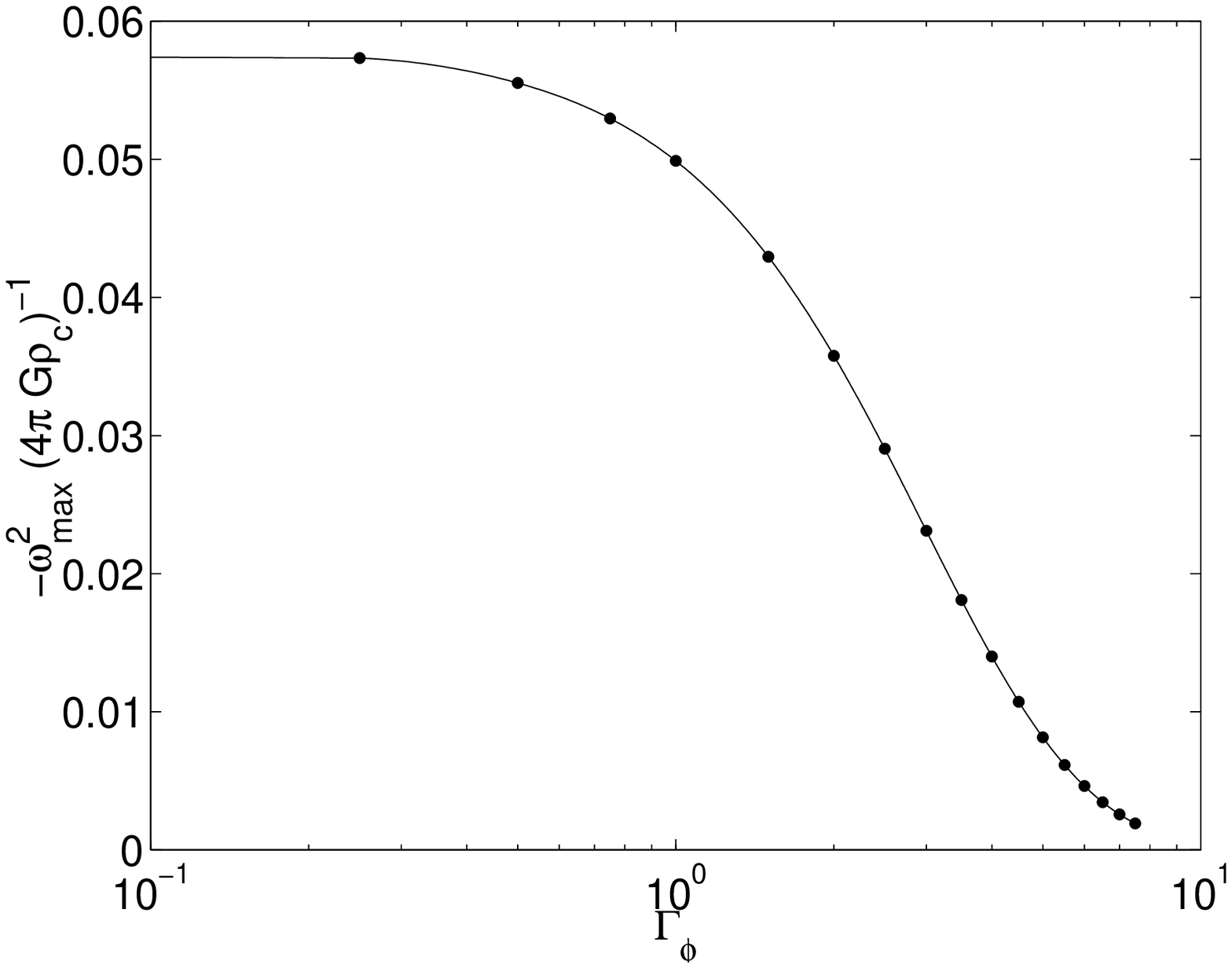,width=\linewidth}  
\caption{We plot the wave numbers $k_{z,max}$ and $k_{z,crit}$, as well as the squared growth rate $-\omega^2$ of the most
unstable mode, as a function of the $\Gphi$.}
\label{fig:conttor}
\end{figure}

\begin{figure}
\begin{minipage}{.49\linewidth}
\epsfig{file=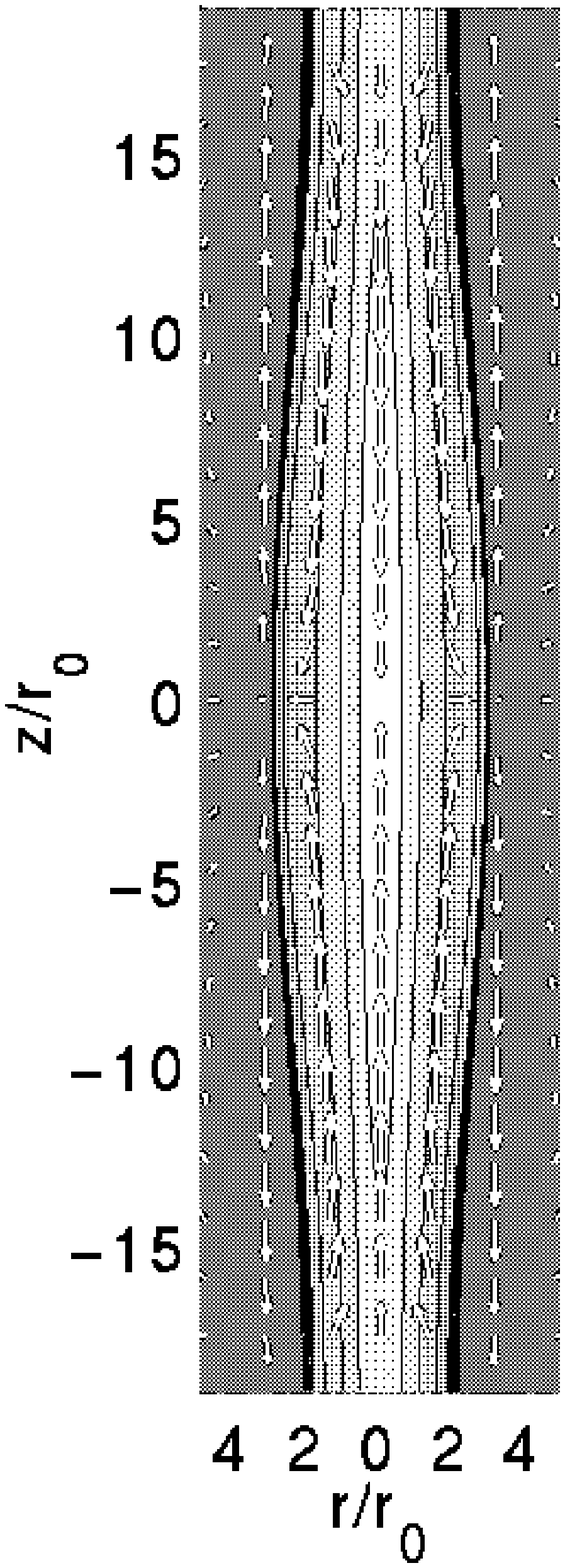,width=\linewidth}
\end{minipage}
\begin{minipage}{.49\linewidth}
\epsfig{file=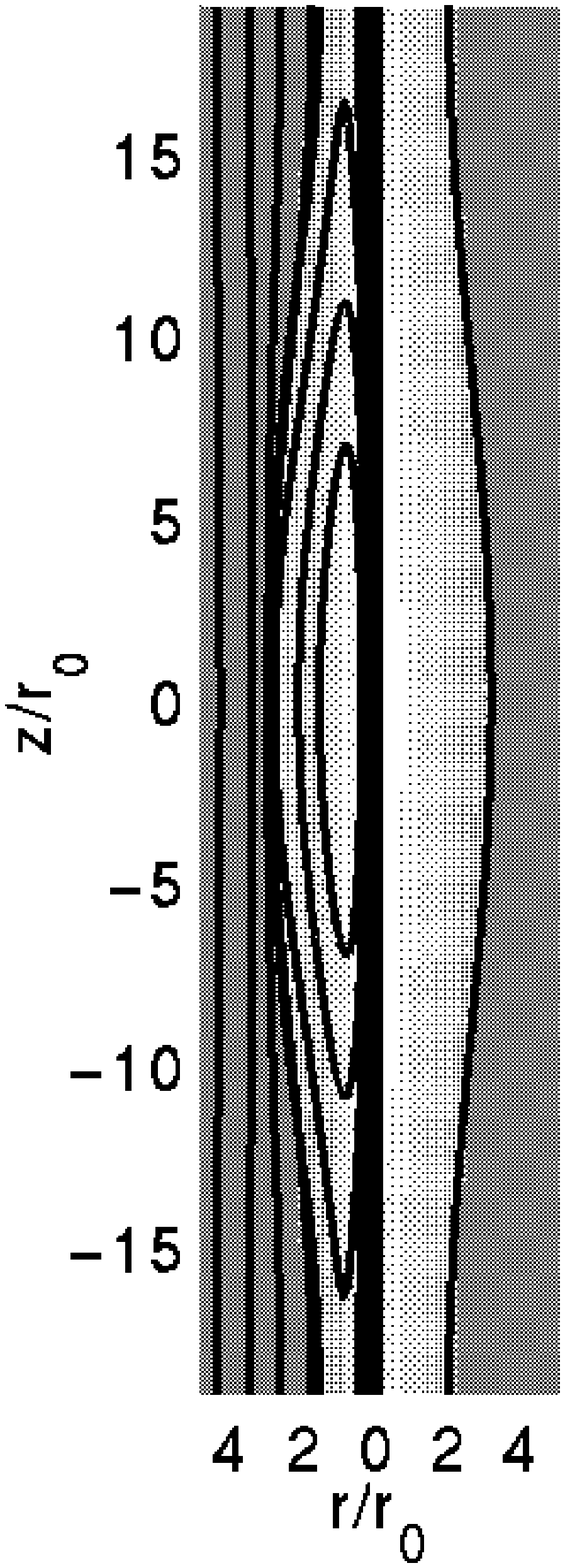,width=\linewidth}
\end{minipage}
\caption{We show the eigenmode corresponding to the most unstable mode
of a truncated magnetized model with a toroidal field ($\Gphi=5$), and mass per unit length
$m/m_0=5$ ($m/\mvir=0.199$). For this mode, $k_{z,max}=0.165$ and $-\omega^2 (4\pi G\rho_c)^{-1}=0.0081$.  
The figures represent a) (left panel) density contours with superimposed poloidal velocity vectors.  
We note that there is no toroidal velocity when the field is purely 
toroidal. and b) (right panel) The magnetic field.  We always represent the magnetic field as
a split-frame figure, with poloidal field vectors shown on the right, and contours to represent the toroidal field on the left.  Since no poloidal
field is generated in this case, the right side of the figure is left blank.}
\label{fig:toreig}
\end{figure}

Finally we consider models in which the magnetic field is purely toroidal.  We stress that these models are illustrative only,
and certainly do not represent realistic models of filamentary clouds.  Figure \ref{fig:tor} shows a sequence
of dispersion curves, in which we have varied the toroidal flux to mass ratio $\Gphi$ while holding the 
mass per unit length constant at $m=5~m_0$ ($m/\mvir=0.199$).
We observe that the toroidal field strongly stabilizes the cloud against fragmentation.  Contrary to the effects
of the poloidal field, the stabilization does not saturate; in fact, the models are completely stable against
fragmentation when $\Gphi \appgeq 10$.  Purely toroidal fields have a stabilizing effect because fragmentation
requires a flow of material along the axis of the filament, which must result in substantial compression of the toroidal flux tubes.
This results in a gradient in the magnetic pressure $1/8\pi ~\partial B_{\phi}^2/\partial z$ which 
resists this motion.
When the toroidal flux to mass ratio $\Gphi$ is sufficiently high, the toroidal flux tubes can resist compression and
arrest fragmentation altogether.  The effects of the varying $\Gphi$ are most readily apparent in Figure \ref{fig:conttor},
where we have plotted $k_{z,max}$, $k_{z,crit}$, and $-\omega_{max}^2$ as a function of of $\Gphi$.  We have shown
an example of an eigenmode in figures \ref{fig:toreig}, where $\Gphi=5$.
In both cases, the toroidal field becomes strongest near the fragments.
The gas motions are purely poloidal in both cases, with a ``reverse flow'' just outside of the filament.  As in the purely
hydrodynamic case, the gas is pushed radially outward near the fragment, and flows toward the empty regions between them.
However, the pinch of the toroidal field restricts the motions to a narrow band just outside the filament.

\begin{table}
\begin{tabular}{l|c|c|c|c|c|c}
\hline
& $\Gphi$ & C & $P_S/\Pave$ & $\tilde{k}_{z,max}$ & $-\tilde{\omega}_{max}^2 $ & $\tilde{k}_{z,crit}$ \\
\hline
1   & 0   & 0.149   & 0.801    & 0.462   & 0.0581    & 0.846 \\ 
2   & 1   & 0.162   & 0.764    & 0.427   & 0.0499    & 0.787 \\
3   & 2   & 0.199   & 0.683    & 0.358   & 0.0357    & 0.655 \\
4   & 3   & 0.253   & 0.598    & 0.284   & 0.023     & 0.513 \\
5   & 4   & 0.318   & 0.523    & 0.218   & 0.0139    & 0.39 \\
6   & 5   & 0.392   & 0.462    & 0.165   & 0.00811   & 0.293 \\
7   & 6   & 0.471   & 0.413    & 0.123   & 0.0046    & 0.217 \\
8   & 7   & 0.555   & 0.372    & 0.0914  & 0.00256   & 0.16  \\
\hline
\end{tabular}
\caption{We give the the wave numbers and growth rates for the fastest growing modes of
truncated filaments with toroidal field only (See Figure \ref{fig:tor}).
We also give the critical growth rate, beyond which the filament is stable against axisymmetric
perturbations.  In all cases,
the mass per unit length is $5 m_0$ ($m/\mvir=0.199$), which falls within the range of observationally allowed
masses per unit length found in Paper I.  We have also assumed an external medium which is perfectly conducting
and of finite density, with $\sigma_e=5 \sigma$.}
\label{tab:fig3}
\end{table}

\section{Stability of Filamentary Clouds With Helical Magnetic Fields}
\label{sec:helix}
In this Section, we consider the stability of models that best agree with the observational constraints found 
in Paper I:
\bea
0.11 \leq \frac{m}{\mvir} \leq 0.43 \nn\\
0.012 \leq \frac{\Ps}{\Pave} \leq 0.75. \nn\\
0.2 \leq X \leq 5, \\
\label{eq:constraint1}
\eea
where the reader is refered to Paper I for a description of the various quantities.
We used a Monte Carlo exploration of our three dimensional ($\Gz$, $\Gphi$, and $C$)
parameter space to find a set of models that agree with these constraints.  We use the same set of models
here to determine the stability properties of filamentary clouds threaded by helical fields.
It is not necessary to show full dispersion curves for all models, since most of the useful information
is contained in the dispersion parameters $k_{z,max}$, $k_{z,crit}$, and $-\omega_{max}^2$.  Thus, we try to 
determine which combination of model parameters $\Gz$, $\Gphi$, and $C$ \, controls the stability properties.

\begin{figure}
\epsfig{file=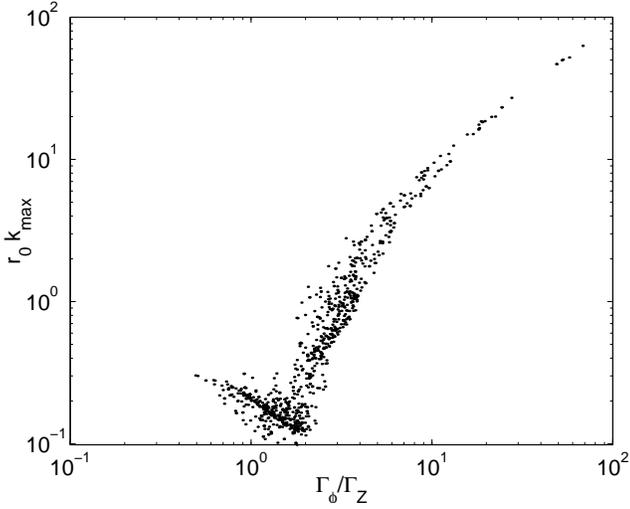,width=\linewidth}
\caption{The most unstable wave numbers for a random sample of models that agree with our constraints.}  
\label{fig:SCAT1}
\end{figure}

\begin{figure}
\epsfig{file=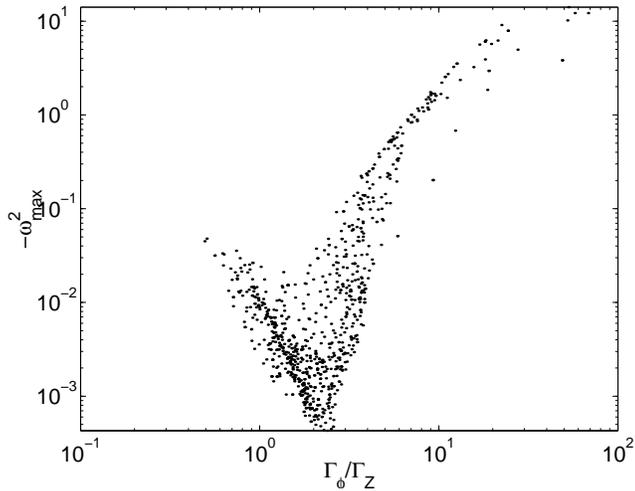,width=\linewidth}
\caption{The growth rates for a random sample of models that agree with our constraints.}
\label{fig:SCAT2}
\end{figure}

\begin{figure}
\epsfig{file=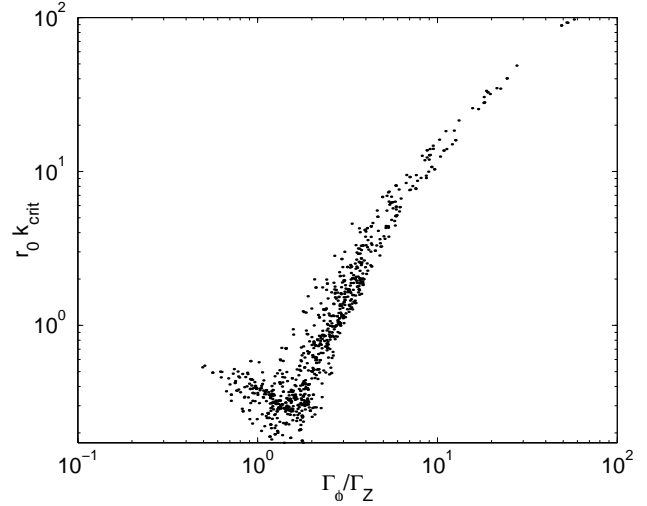,width=\linewidth}
\caption{The critical wave numbers for a random sample of models that agree with our constraints.}
\label{fig:SCAT3}
\end{figure}

We find that the dispersion parameters $k_{z,max}$, $k_{z,crit}$, and $-\omega_{max}^2$ are best correlated with the ratio $\Gphi/\Gz$.
Figures \ref{fig:SCAT1} to \ref{fig:SCAT3} show scatter plots of these dispersion parameters as functions of $\Gphi/\Gz$.
It is obvious from these figures that we have found two types of unstable mode in our calculation; the two types of unstable mode 
are mostly well separated in Figures \ref{fig:SCAT1} to \ref{fig:SCAT3}, but join
together when $\Gphi/\Gz\approx 2$.  The first, which occurs when
$\Gphi/\Gz$ is small, is a gravity-driven instability, which is analogous to the modes found for purely 
hydrodynamic models (Section \ref{sec:ost}),
and MHD models in which the field is purely poloidal or toroidal (Section \ref{sec:pure}).  
We note that increasing $\Gphi/\Gz$ substantially decreases the growth 
rate of the instability.  This is in accord with the findings of Section \ref{sec:pure}, where we found that increasing $\Gphi$ and $\Gz$
both stabilize gravity driven modes, but $\Gphi$ is much more effective.  In fact, virtually all of the gravity-driven modes in Figure
\ref{fig:SCAT2} have growth rates that are significantly lower than the growth rates of unmagnetized filaments, as well as filaments with purely  
poloidal fields (shown in figures \ref{fig:Ost} and \ref{fig:pol} respectively).
Typically, we find
\bea
0.0005 &\appleq & -{\tilde\omega}_{max}^2 \appleq 0.03 \nn\\
0.1 &\appleq & \tilde{k}_{z,max} \appleq 0.3 \nn\\
0.2 &\appleq & \tilde{k}_{z,crit} \appleq 0.6, 
\eea
where the tildes refers to the dimensionless growth rates and wave numbers, as defined
in equation \ref{eq:dimensionless}.

By taking the inverse of $\sqrt{-\omega^2}$, we find a growth timescale of 
\be
\tau_{frag} = 1.8 \left(\frac{-\tilde{\omega}_{max}^2}{0.01}\right)^{-1/2}\left(\frac{n_c}{10^4~cm^{-3}}\right)^{-1/2}~Myr.
\label{eq:tauf}
\ee
where $n_c$ is the central number density of the filament.
We have chosen a fiducial central density of $10^4~cm^{-3}$ because radially extended
($0.5~pc$ diameter) filamentary structure is clearly visible
in $C^{18}O$ maps of Taurus (Onishi et al. 1998).  $C^{18}O$ molecules require a density of at least
$\sim 2\times 10^3~cm^{-3}$ for excitation, and our models predict that the central density should be
several times higher than the bulk of the filament.
Nevertheless, the central densities of filaments have not yet been accurately measured, 
so this fiducial central density should be treated with 
caution.  The radial signal crossing time is approximately given by
\be
\tau_x\approx\frac{Rs}{\sigma}=0.49\left(\frac{\Rs}{0.25~pc}\right)\left(\frac{0.5~km~s^{-1}}{\sigma}\right)~Myr.
\label{eq:taux}
\ee
Since the signal crossing time is comparable to, and probably slightly longer than the fragmentation timescale,
we expect filamentary clouds to achieve radial quasi-equilibrium before fragmentation destroys the filament in a few
times $\tau_{frag}$; thus, our analysis is consistent
with the assumption of quasi-equilibrium used in Paper I.

The second type of unstable mode is driven by the magnetic field.
These MHD-driven instabilities are triggered when $\Gphi/\Gz \appgeq 2$.
Figures \ref{fig:SCAT1} to \ref{fig:SCAT3} show that this mode joins onto the gravity-driven modes, but extends to very high growth rates and very large wave numbers.
Since large wave numbers correspond to small wavelengths, gravity cannot be important, as a driving force, for these modes.
They are, however, most unstable when the toroidal flux wrapping the filament is large, compared to the poloidal flux.  This suggests that the mode is an MHD instability.
In fact the criterion that we have found is analogous to the famous stability criterion for MHD ``sausage'' modes: 
\be
\Bphis > \sqrt{2} \Bz
\label{eq:sausage}
\ee
(cf. Jackson 1975).  The analogy is far from perfect because equation \ref{eq:sausage} applies to the case of a uniform plasma
cylinder threaded by a constant poloidal magnetic field, and surrounded by a toroidal field driven by a thin axial current sheet
along the surface of the plasma.  However, the axial current distribution in our case is distributed throughout the filament.
Nevertheless, the similarity between equation \ref{eq:sausage} and our own instability criterion
is very suggestive.  

We do not find MHD-driven instabilities for models with purely toroidal fields.  We note that the stability criterion
for a non-self-gravitating plasma column pinched by a purely toroidal field can be expressed analytically as
\be
-\frac{d \ln P}{d \ln r} < \frac{4}{2+\beta}
\label{eq:stability}
\ee
for an isothermal perturbation, where $\beta$ is the usual plasma $\beta$, defined by $\beta=8\pi P/\Bphi^2$
(cf. Sitenko \& Malnev 1995).  We have checked, numerically, that our equilibria obey equation \ref{eq:stability}, 
and, therefore, should be stable against ``sausage'' modes.  It may appear that equation \ref{eq:sausage} is violated 
for any model with a purely toroidal field, but we remind the reader that equation \ref{eq:sausage} strictly 
applies to situations in which the toroidal field is generated by a current sheet at the plasma surface, and thus 
resides entirely outside of the plasma column.  Equation \ref{eq:stability} is, however, appropriate for the 
more distributed fields in our models.

It seems paradoxical that models with purely toroidal fields are stable against ``sausage'' modes,
while those with a poloidal field component may be unstable.  However, we note that helical fields
are topologically different from purely toroidal fields.  Flux tubes in purely toroidal models
form closed loops, while the loops of toroidal flux are all linked in the case of a helical field.
When the field is helical, twisting motions in the filament may enhance the toroidal field, while no amount of
twisting can amplify a purely toroidal field.  Thus, we argue that it is at least possible for modes to exist
that simultaneously generate twisting motions and toroidal field, which could destabilize the filament.

We note that the gravity-driven modes join continuously with the
MHD-driven modes in Figures \ref{fig:SCAT1} to \ref{fig:SCAT3}; the
intersection of the two branches, when $\Gphi/\Gz\approx 2$, represents a transition between gravity
and the toroidally dominated magnetic field as the main driving force of the instabilty.
Some of the modes that we have labeled as magnetic have growth
rates that are similar to gravity modes, except that the wavelength is
somewhat shorter.  These may represent reasonable modes of fragmentation
for filamentary clouds.  However, the more unstable modes
of the MHD branch probably cannot, since they grow too rapidly and would probably disrupt the filament
on timescales that are much shorter than the timescale on which equilibrium can be established
(See equations \ref{eq:tauf} and \ref{eq:taux}).  Therefore,
a subset of the models allowed by the observational constraints of FP1
are actually highly unstable.

\begin{figure*}
\begin{minipage}{.32\linewidth}
\epsfig{file=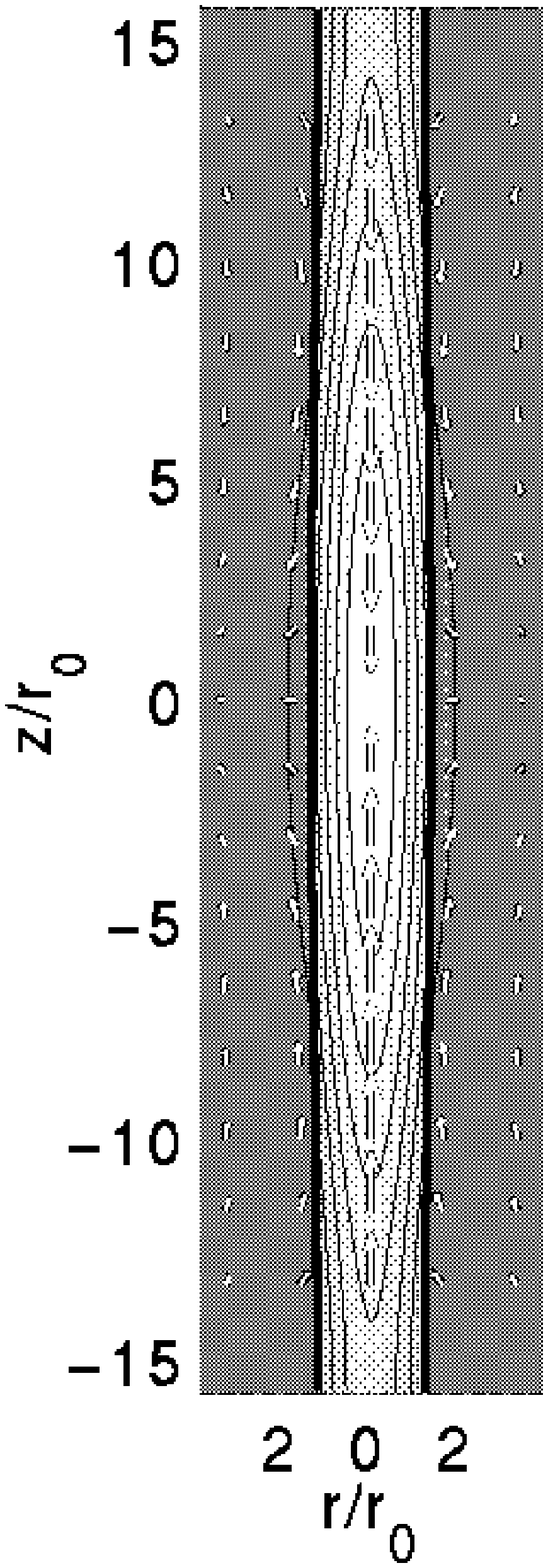,width=\linewidth}
\end{minipage}
\begin{minipage}{.32\linewidth}
\epsfig{file=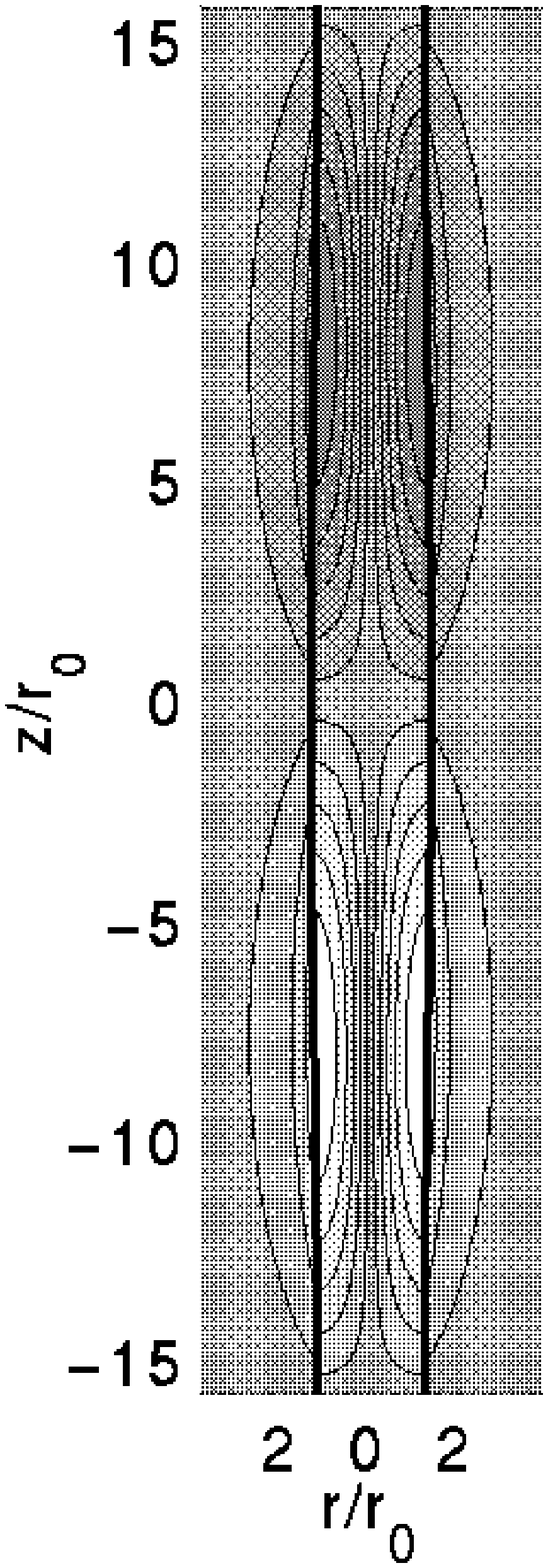,width=\linewidth}
\end{minipage}
\begin{minipage}{.32\linewidth} 
\epsfig{file=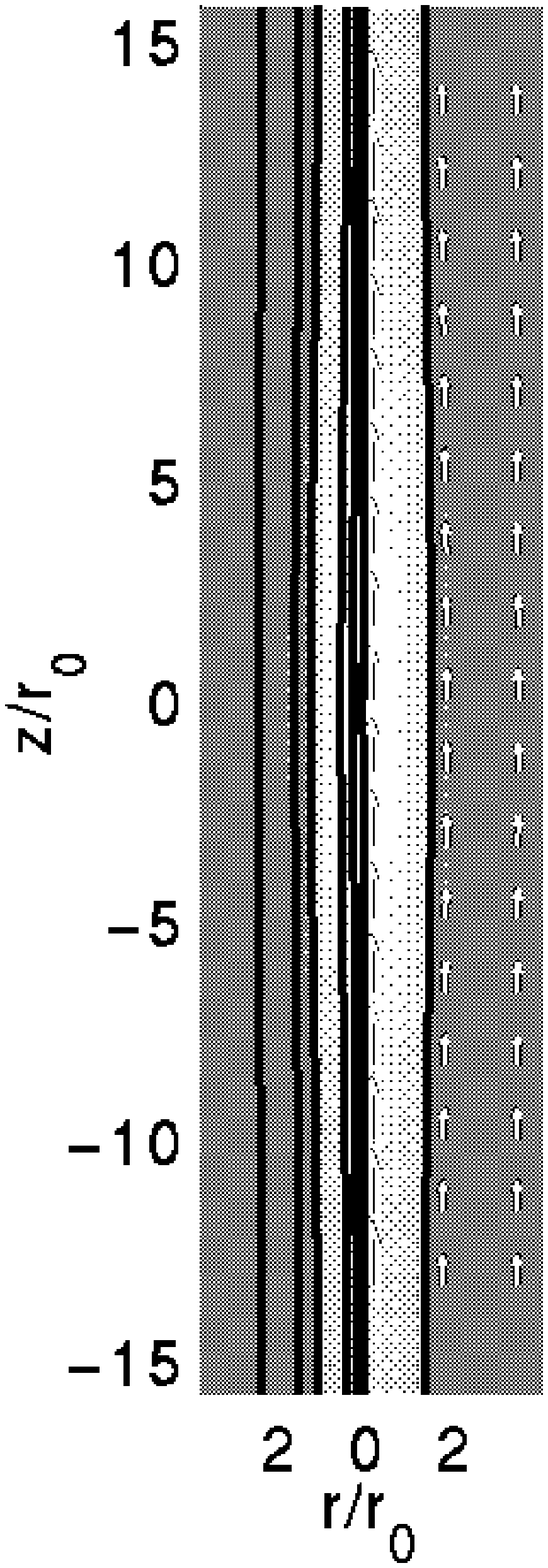,width=\linewidth}
\end{minipage}
\caption{An example of a gravity-driven mode for a model with a helical magnetic field.
For this mode, $k_{z,max}=0.199$ and $-\omega^2 (4\pi G\rho_c)^{-1}=0.0081$.  
The figures represent a) (left panel) density
contours with superimposed poloidal velocity vectors.  b) (middle panel) Contours of toroidal velocity.  Light coloured contours
represent positive $v_\phi$, while dark coloured contours represent negative $v_\phi$.
and c) (right panel) The magnetic field.  We always represent the magnetic field as
a split-frame figure, with poloidal field vectors shown on the right, and contours to represent the toroidal field on the left.}
\label{fig:helixgrav}
\end{figure*}

\begin{figure}
\begin{minipage}{\linewidth}
\epsfig{file=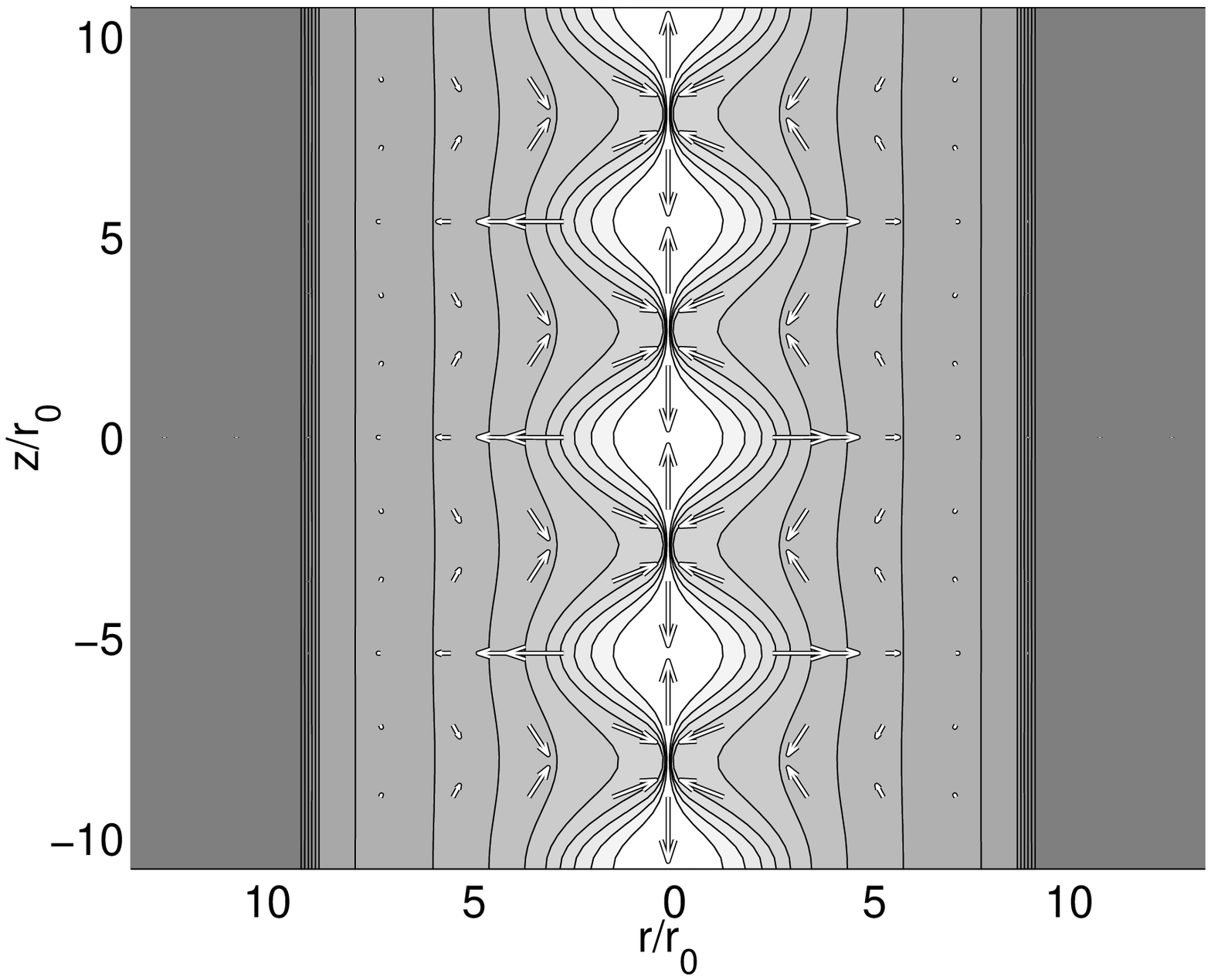,width=.9\linewidth}
\end{minipage}

\begin{minipage}{\linewidth}
\epsfig{file=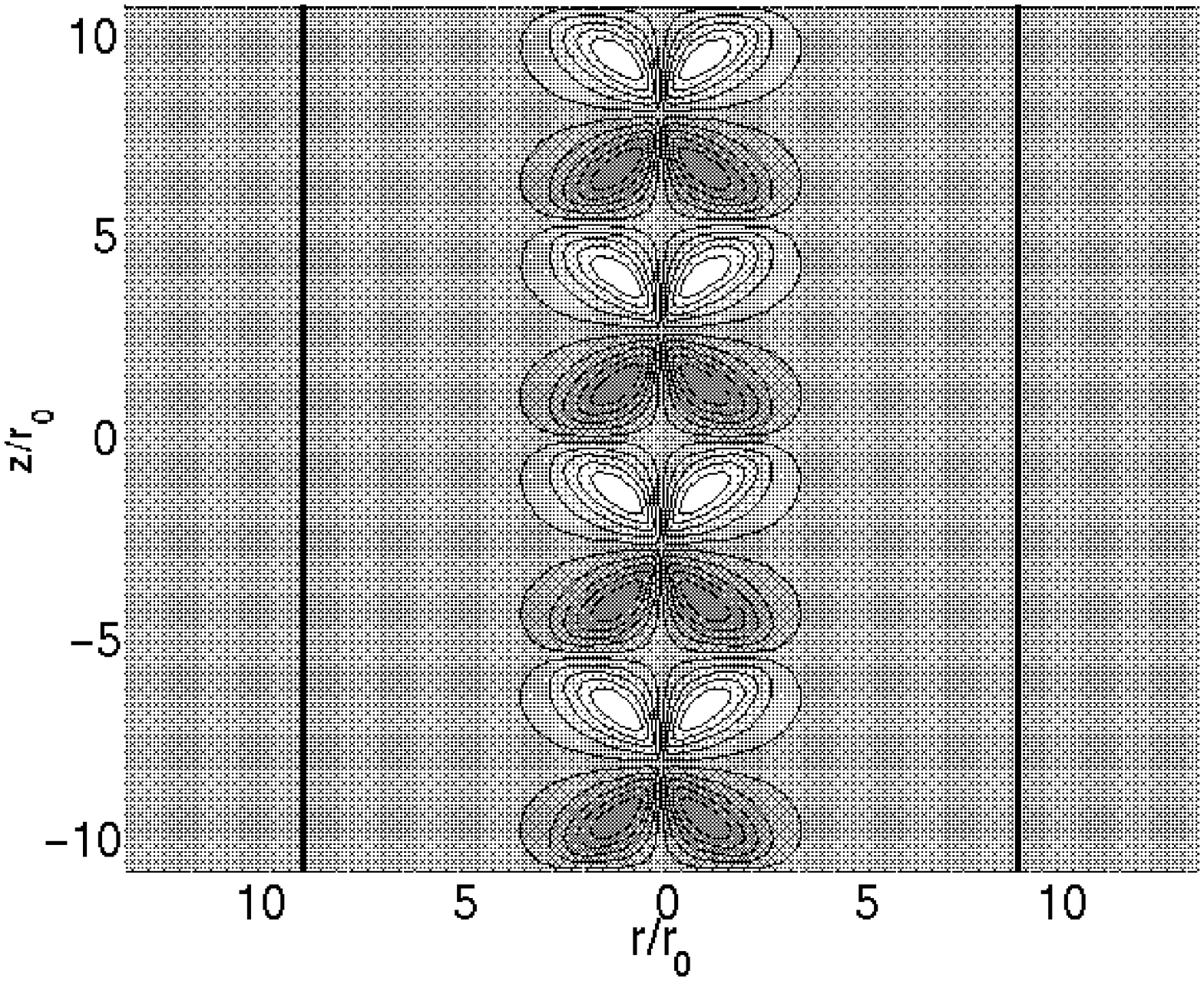,width=.9\linewidth}
\end{minipage}

\begin{minipage}{\linewidth}
\epsfig{file=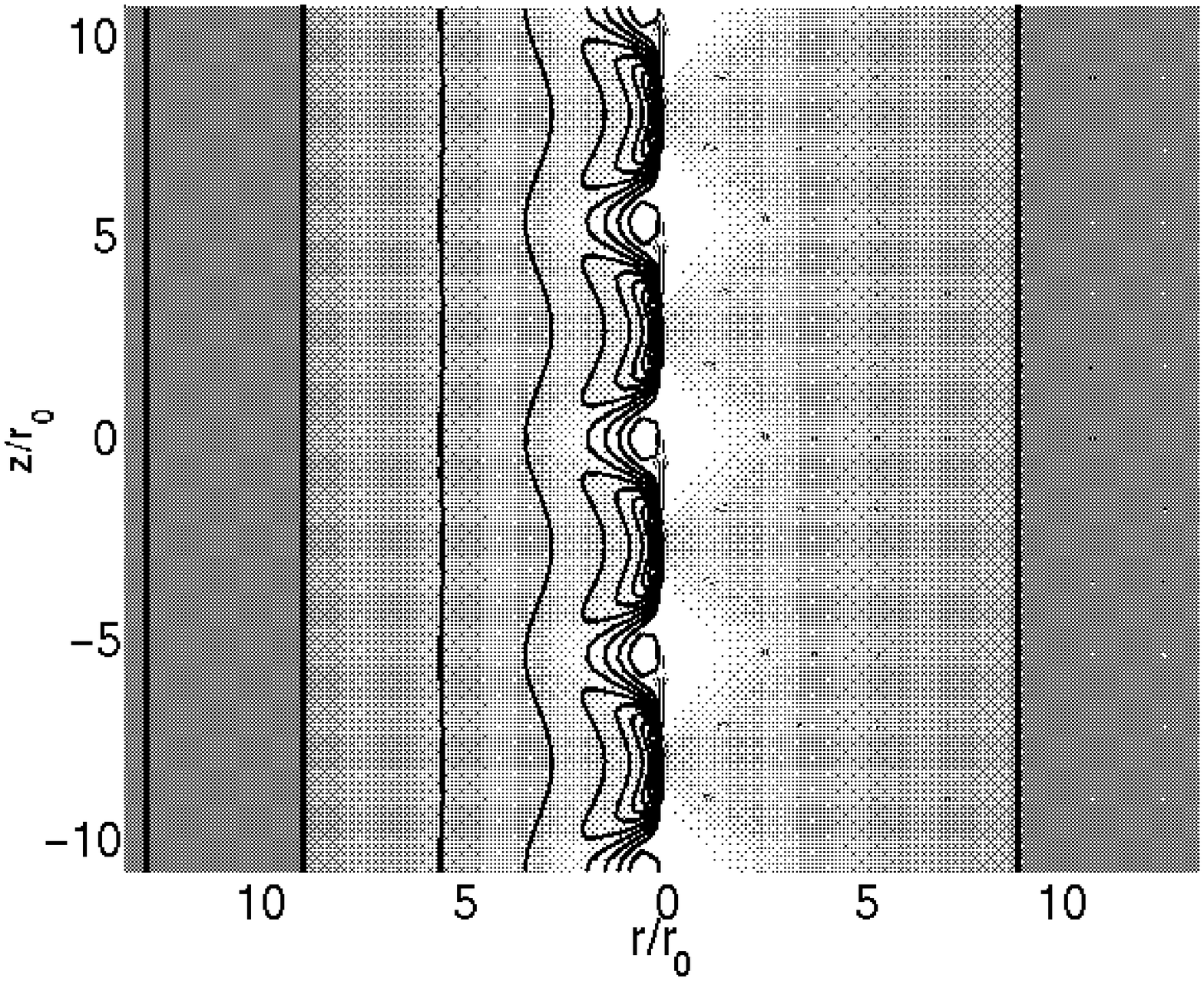,width=.9\linewidth}
\end{minipage}
\caption{An example of an MHD-driven mode for a model with a helical magnetic field.
For this mode, $k_{z,max}=1.17$ and $-\omega^2 (4\pi G\rho_c)^{-1}=0.0390$.  
The figures represent a) (top panel) density
contours with superimposed poloidal velocity vectors.  b) (middle panel) Contours of toroidal velocity.  Light coloured contours          
represent positive $v_\phi$, while dark coloured contours represent negative $v_\phi$.
and c) (bottom panel) The magnetic field.  We always represent the magnetic field as
a split-frame figure, with poloidal field vectors shown on the right, and contours to represent the toroidal field on the left.}                 
\label{fig:helixmhd}
\end{figure}

In Figure \ref{fig:helixgrav}, we show an example of an unstable gravity-driven mode.  
As in the previously considered cases of hydrodynamic filaments (Section \ref{sec:ost}) and filaments with
only one field component (Section \ref{sec:pure}), gravity
drives a purely poloidal flow of gas towards the fragments.  
However, a helical field results in a toroidal component of the velocity field which alternates in sign
from one side of the fragment to the other.  The reason is that the helical field
exerts oppositely directed torques on the gas flowing towards the fragment from opposite directions. 
The mode shown in Figure \ref{fig:helixmhd} is driven by the MHD
``sausage'' instability.  The dominant wavelength of the instability is much shorter than either the filament radius or the dominant wavelength
of the gravity-driven modes; therefore gravity is relatively unimportant
compared to the magnetic stresses.  The actual structure of the mode
is quite similar to that of the gravity-driven modes except that the
mode is confined to the most central parts of the filament, where
the $\Bphi$ gradient is steepest.  This is consistent with our claim
that this mode is a ``sausage'' instability, because ``sausage'' modes
are driven by outwardly increasing toroidal fields with strong 
gradients.  The filament is crushed where the toroidal
field is strongest, which forces gas out along the axis and
towards the fragments.  
As in the case of the gravity-driven modes, toroidal
motions are generated as the helical field 
exerts torques on the gas.

\begin{figure}
\epsfig{file=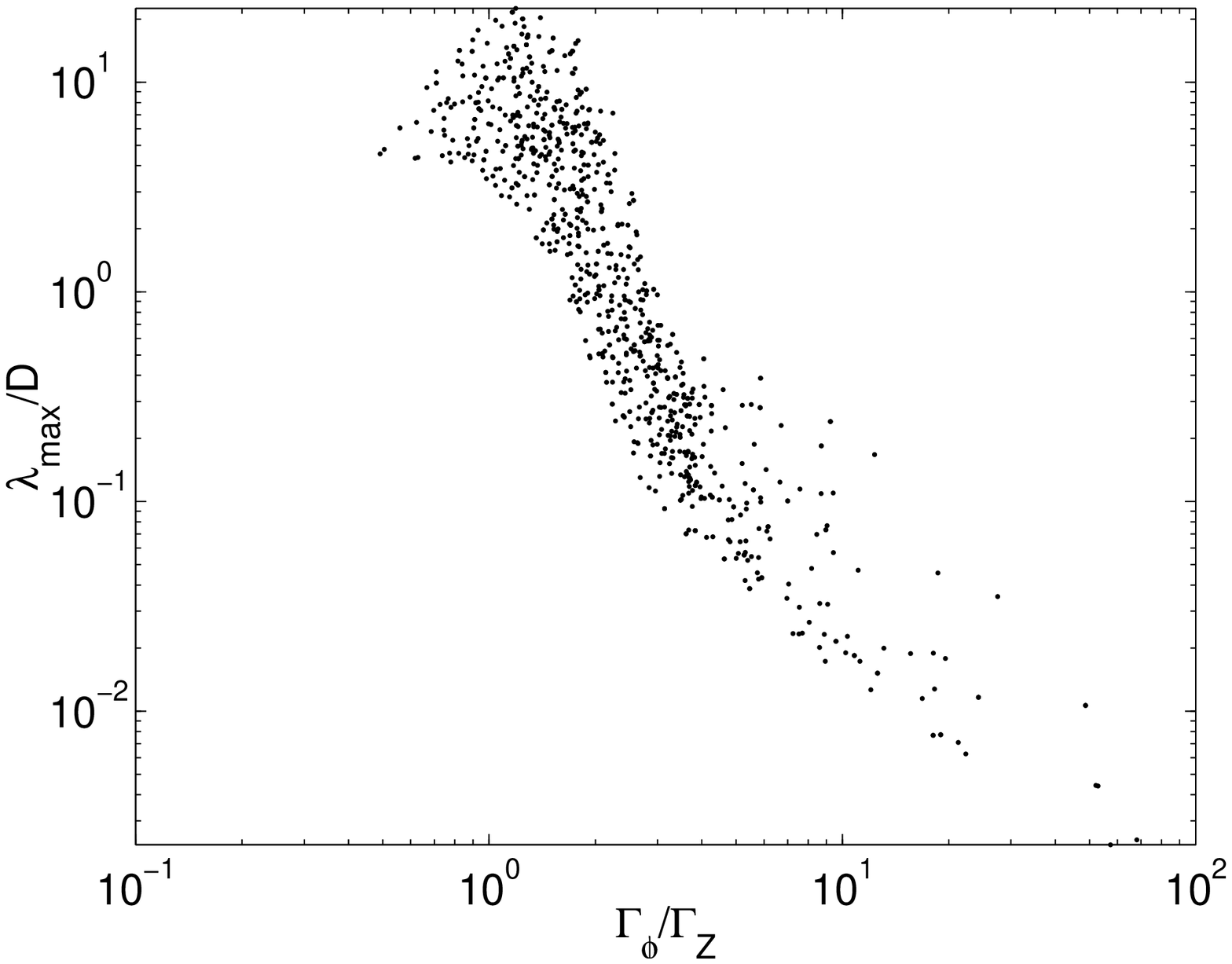,width=\linewidth}
\caption{The ratio of fragment separation $\lambda_{max}=2\pi/k_{max}$ to filament diameter $D=2\Rs$
for a random sample of models that agree with our constraints.}
\label{fig:SCAT4}
\end{figure}

Figure \ref{fig:SCAT4} shows the effect of $\Gphi/\Gz$ on the ratio
of the fragmentation wavelength $\lambda_{max}$
to the filament diameter $D$, which we define as twice the filament radius $\Rs$; 
we note that gravity-driven and MHD-driven modes cannot be
easily separated on this diagram.
In all cases, we find that more toroidally dominated modes
have smaller $\lambda_{max}/D$.  The Schneider and Elmegreen (1979) Catalogue
of Dark Globular Filaments shows that $\lambda_{max}/D \approx 3.0$ for most
filaments in their sample. Interestingly, this value is consistent with $\Gphi/\Gz\approx 2$,
which is near the transition from gravity to MHD-driven modes, and is near the lowest possible 
growth rate.  We postulate that most filamentary clouds may be observed to reside near this maximally stable 
configuration because such objects would survive the longest.  Although suggestive, this conclusion will 
also depend on the non-axisymmetric stability of filamentary clouds, to be discussed in a forthcoming paper.

\begin{figure}
\epsfig{file=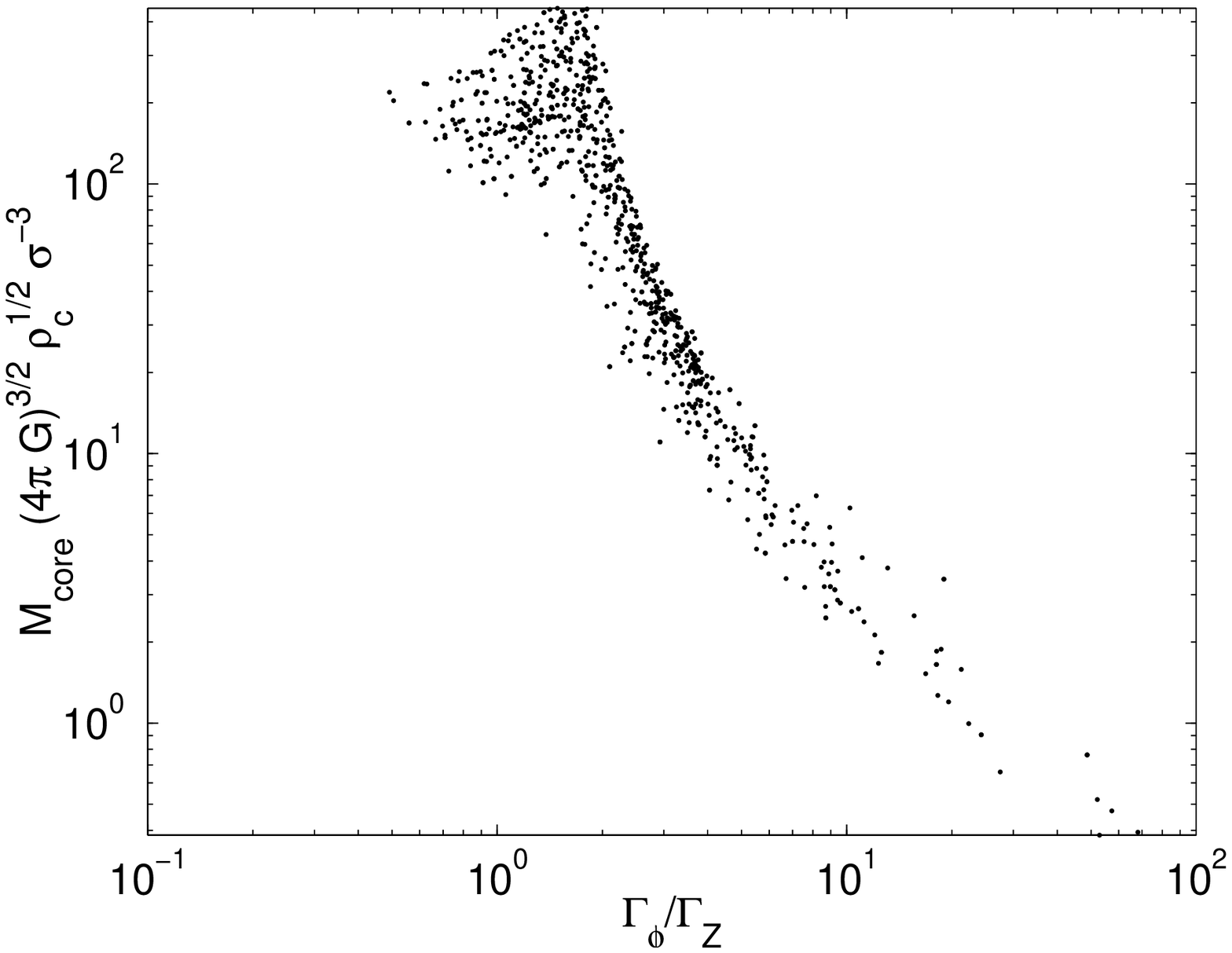,width=\linewidth}
\caption{The expected mass of fragments for a random sample of models that agree with our constraints.
Here we assume that $M_{frag}=m \lambda_{max}$.}
\label{fig:SCAT5}
\end{figure}

Finally, in Figure \ref{fig:SCAT5}, we show the masses of 
fragments that might form from filamentary clouds.  We note that we have computed the fragment masses simply 
by multiplying the equilibrium mass per unit length by the wavelength of the most unstable mode.
This procedure implicitly assumes that there is no sub-fragmentation as fragments evolve.  
Thus, these masses should be regarded
with caution.  We find that fragment masses, computed in this way, fall in a rather large range; in dimensionless
units we find
\be
100 \appleq M_{frag} (4\pi G)^{3/2} \rho_c^{1/2} \sigma^{-3} \appleq 400
\ee
for gravity-driven modes, which have $\Gphi/\Gz\appleq 2$.
Thus, we obtain masses of
\be
M_{frag}=41.5 \left(\frac{\tilde{M}_{frag}}{100}\right) \left(\frac{\sigma}{0.5 km~s^{-1}}\right)^3 \left(\frac{n_c}{10^4~cm^{-3}}\right)^{-1/2} M_\odot.
\ee 

\section{Discussion and Summary}
\label{sec:discussion}
We have determined how pressure truncation and helical fields affect the stability 
of our models against axisymmetric modes of fragmentation,
which are ultimately responsible for the formation of cores.
Our analysis differs from previous work on filaments in a number of respects.  Firstly, and most importantly,
we have made at least a first attempt to obervationally constrain our models; thus, our modes of
fragmentation pertain to equilibria that are truncated by realistic external pressures and
have magnetic fields that are consistent with the observations.  Moreover, our models have
density gradients of $r^{-1.8}$ to $r^{-2}$, in good agreement with observational data (Alves et al 1998, Lada et al. 1998).
Secondly, we have treated the external medium self-consistently by considering it to be a perfectly conducting gas of finite density.

We found, in Paper I, that filaments that are highly truncated by the external pressure 
have lower masses per unit length than untruncated filaments.
We have shown, in Section \ref{sec:ost},
that such pressure dominated equilibria are more stable than their less truncated gravitationally
dominated counterparts.
We also found that the finite density of the external medium, in our calculation, slightly
stabilizes filaments due to inertial effects.  
This effect vanishes when the external medium is ``hot'', with a
velocity dispersion $\appgeq 10 ~\sigma$.

Both the poloidal and toroidal fields were shown to stabilize filaments against gravity-driven
fragmentation in Section \ref{sec:pure}.
The poloidal field stabilizes the filament by resisting motions perpendicular to the poloidal field lines.
This stabilization eventually saturates, when the field is strong enough ($\Gz\appgeq 10$) to disallow radial motions altogether.  
Surprisingly, the toroidal field is
much more effective at stabilizing filaments against gravitational fragmentation.  As the filaments begins to fragment, the 
toroidal flux tubes are pushed together near the fragments; this results in a gradient in the magnetic pressure
($1/8\pi ~\partial B_{phi}^2/ \partial z$)
that opposes fragmentation.  In the unrealistic limit of purely toroidal fields, the filament is stable
to fragmentation when $\Gphi \appgeq 7$.  One might expect that these models would be highly unstable against
MHD ``sausage'' modes.  However, we have verified that our equilibria satisfy the stability condition given in equation
\ref{eq:stability}.

In Section \ref{sec:helix}, we demonstrated that our models with helical fields are subject to
both gravity-driven and MHD-driven instabilities.
The two types of modes blend together in Figures \ref{fig:SCAT1} to \ref{fig:SCAT3}, with the transition
from gravity-driven to MHD-driven largely determined by the ratio $\Gphi/\Gz$; when $\Gphi/\Gz\appleq 2$, the instability
is driven by gravity, with MHD-driven modes triggered only for more tightly wrapped helical fields.
We provided evidence, in Section \ref{sec:helix}, that the MHD-driven mode is probably is an MHD ``sausage'' mode.
These instabilities can have extremely high growth rates compared to gravity driven modes; therefore, we may interpret our stability 
condition as an additional theoretical constraint on our models. 
We note that the most stable of all equilibria are near the transition from grivity-driven to
MHD-driven modes.  Thus, we find that there is a regime
in which filamentary clouds fragment very slowly.
For gravity-driven modes, we find that $-\omega_{max}^2$, $k_{z,max}$, and $k_{z,crit}$ fall within the 
ranges
\bea
0.0005 &\appleq & -\omega_{max}^2 (4\pi G \rho_c)^{-1} \appleq 0.03 \nn\\
0.1 &\appleq & r_0 k_{z,max} \appleq 0.3 \nn\\
0.2 &\appleq & r_0 k_{z,crit} \appleq 0.6,
\eea
where $r_0$ is the core radius for filamentary clouds (see Paper I), and $\rho_c$ is the central density.
We may also write the expected wavelength for the separation of fragments and the growth timesacale
as
\bea
\lambda_{max} &=& 2.8 \left(\frac{\tilde{k}_{max}}{0.2}\right)^{-1}\left(\frac{\sigma_c}{0.5~km~s^{-1}}\right)
	\left(\frac{n_c}{10^4~cm^{-3}}\right)^{-1/2}~pc \nn\\
\tau_{frag} &=& 1.8 \left(\frac{-\tilde{\omega}_{max}^2}{0.01}\right)^{-1/2}\left(\frac{n_c}{10^4~cm^{-3}}\right)^{-1/2}~Myr.
\eea
We note that the growth time for gravity-driven modes is much longer than the radial
signal crossing time (equation \ref{eq:taux}), on which radial equilibrium is established.  Therefore, our stability analysis is
consistent with the assumption of radial quasi-equilibrium used in Paper I.

We find that the fragmentation of filamentary clouds with helical fields leads to a toroidal
component of the velocity field, whose direction alternates with every half wavelength of the perturbation.
These motions are comparable to, and in some cases exceed the poloidal velocities, at least in linear theory.  
These motions could, in principal, be detected, which would provide considerable evidence for helical fields.

We find that the fiducial growth time for gravity-driven modes is on the order of 1.8 Myr, which is longer than the 
signal crossing time, on which radial equilibrium is established.  Therefore, our stability analysis is 
consistent with the assumption of radial quasi-equilibrium used in Paper I.

Throughout this analysis, we have focussed on axisymmetric modes that lead to fragmentation,
ignoring any non-axisymmetric modes that might be present.  Some of our models may 
be unstable to at least the $m=1$ ``kink'' instability.  However, we do not expect gross instability for 
most models.  While they do contain significant toroidal fields, the poloidal field 
is actually much stronger throughout most of the filament, particularly in the central regions,
where $\Bz$ is maximal and $\Bphi$ vanishes.  This ``backbone'' of poloidal field should largely
stabilize our models against kink modes.  In any case, the non-axisymmetric modes of
fragmentation will be addressed in a forthcoming paper.

\section{Acknowledgements}
J.D.F. acknowledges the financial support of McMaster University and an Ontario Graduate Scholarship.
The research grant of R.E.P. is supported by a grant from the Natural Sciences and Engineering
Research Council of Canada.

\appendix
\section{The Equations of Linearized MHD}
\label{app:MHD}
The equations of linearized, self-gravitating, perfect MHD that we solve are as follows:\\
Momentum Equation:
\bea
i\omega\rho_0\vi &+& \nabla P_1 + \rho_0\nabla\Phi_1+\rho_1\nabla\Phi_0 \nn\\
	         &-&\frac{1}{4\pi}\left[(\nabla\cross\Bo)\times\Bi + (\nabla\cross\Bi)\times\Bo\right]=0 \nn\\ 
\label{eq:dyn}   
\eea
Mass conservation:
\be
i\omega\rho_1+\nabla\cdot(\rho_0\vi)=0
\label{eq:masscons}  
\ee
Induction equation:
\be
i\omega\Bi=\nabla\cross(\vi\cross\Bo).
\label{eq:induction}
\ee
Poisson's equation:
\be
\nabla^2\Phi_1=\rho_1.
\label{eq:poisson}
\ee
Equation of state:
\be
P_1=\gamma\sigma_0^2\rho_1,
\label{eq:EOS}
\ee
where $\gamma$ is the adiabatic index of the gas ($\gamma=1$ within the filament.)
and $\sigma_0^2=P_0/\rho_0$ is the one-dimensional velocity dispersion.

By introducing a ``modified'' gravitational potential $\varphi_1=i\omega\Phi_1$, it
is straightforward to combine equations \ref{eq:dyn} to \ref{eq:EOS} to 
form the eigensysem given by equations \ref{eq:eig1} and \ref{eq:eig2b}.
Equations \ref{eq:eig1} and \ref{eq:eig2b} represent the eigensystem, written entirely in terms
of $\varphi_1$ and the components of $\rho v_1$.
We now define differential operators ${\hat A}$, ${\hat B}$, ${\hat C}$, and ${\hat D}$
in a way that is obvious from equations \ref{eq:eig1} and \ref{eq:eig2b} to write our
eigensystem in closed symbolic form:
\bea
-\omega^2\rho_0\vi &=& {\hat A} (\rho_0\vi)+ {\hat B} \varphi_1 \nn\\
0 &=& {\hat C} (\rho_0\vi) + {\hat D} \varphi_1. \\
\label{eq:eig3}
\eea

We obtain an approximate matrix representation of operators ${\hat A}$, ${\hat B}$, ${\hat C}$, and ${\hat D}$
by finite differencing over a one-dimensional grid of $N$ cells ($N\approx 500$ usually).  From this point forward, we
shall assume that all operators have been finite differenced, and shall make no distinction between matrix and
operator forms.  We note that equation \ref{eq:eig3} applies to both the HI envelope and the molecular filament; thus, of the $N$ cells,
some portion (usually $\sim 2/3$) are within the filament, while the remainder are in the HI envelope.  Since the density is
discontinuous at the interface, care must be taken not to difference any equations across the boundary.  Section \ref{sec:BCS} describes the
boundary conditions that link the perturbation in the filament to the external medium.  Matrices ${\hat C}$ and ${\hat D}$
are sparse tridiagonal $N\times N$
matrices, while ${\hat A}$ and ${\hat B}$ each take the form of $3\times 3$ blocks of $N\times N$ sparse tridiagonal 
sub-matrices (The $3\times 3$ block structure
occurs because these matrices operate on $\rho_0\vi$, which has 3 vector components.).
If we consider ``eigenvector'' ${\bf \Psi}$ given by equation \ref{eq:Psi}, then equation \ref{eq:eig3} takes the form of 
the standard eigenvalue problem given by equation \ref{eq:EIG}, where
\be
{\hat L}=\left[         \begin{array}{cc}
                        {\hat A} & {\hat B} \\
                        {\hat C} & {\hat D} \\
                        \end{array}
        \right].
\ee

\section{The Boundary Conditions at the Molecular Filament/HI Envelope Interface}
\label{app:BC}
Equations \ref{eq:bc1} to \ref{eq:bc4} give the boundary conditions at the interface between the molecular and atomic gas.  
We note that these jump conditions
apply in a Lagrangian frame that is co-moving with the deformed surface.
Equations \ref{eq:bc1} and \ref{eq:bc2} demand that the gravitational potential and its first derivative
(the gravitational field) must be continuous aross the interface.  Equation \ref{eq:bc3} is the usual condition on the
normal magnetic field component from electromagnetic theory.  It is easily derived from the divergence-free condition of the
magnetic field.  The final condition, equation \ref{eq:bc4} states that the normal component of the total stress is
continuous across the boundary.

We must now transform the boundary conditions (equations \ref{eq:bc1}), into an Eulerian frame appropriate to
our fixed grid.  We assume that the deformed surface is defined by the equation
\be
r=\Rs+\epsilon e^{i(\omega t+m\phi+k_z z)},
\ee
where $\epsilon<<\Rs$ for a small perturbation.
We note that $m=0$ for the axisymmetric modes considered in this paper.  However, we retain $m$
in our equations, since we will examine non-axisymmetric modes in a forthcoming paper.
To first order, the unit normal vector to this surface is given by
\be
\nhat={\bf\hat r}-\epsilon\left[ \frac{i m}{\Rs} {\bf\hat\phi}+i k {\bf\hat z} \right] e^{i(\omega t+m\phi+k_z z)}.
\label{eq:normal}
\ee
Since this is a contact discontinuity, and not a shock, the velocity field must be consistent with the motion
of this surface; thus,
\be
v_{r1}=\frac{\partial r}{\partial t}=i\omega\epsilon e^{i(\omega t+m\phi+k_z z)},
\label{eq:vr1}
\ee
which implies that
\be
[v_{r1}]=0
\label{eq:vr1b}         
\ee
{in the Eulerian frame}.
Solving equation \ref{eq:vr1} for $\epsilon$ and substituting into equation \ref{eq:normal}, we obtain an expression
for $\nhat$ that involves only the radial velocity:
\be 
\nhat={\bf\hat r}-v_{r1}\left[ \frac{m}{\omega\Rs} {\bf\hat\phi} + \frac{k_z}{\omega} {\bf\hat z} \right].
\label{eq:nhat}
\ee

We insert the expression for the normal vector $\nhat$ (equation \ref{eq:nhat}) into equations \ref{eq:bc1} to \ref{eq:bc4},
and evaluate all zeroth order quantities,
at the position of the deformed surface, by a first order Taylor expansion.  
We also make use of equations \ref{eq:dyn} to \ref{eq:EOS} in order to express all quantities in terms of
the momentum density $\varphi$ and the potential $\varphi$.
After some algebra, we obtain the explicit form of the boundary conditions that apply at the
surface of the filament:
\bea
~[\varphi_1] &=& 0 \label{eq:BCF1}\\
~\left[\frac{\partial\varphi_1}{\partial r}+\rho_0 v_r \right] &=& 0\label{eq:BCF2}.\\
\left[\frac{1}{\rho_0}(\rho_0 v_r)\right]=0. \label{eq:BCF3} \\
\left[-\gamma\sigma_0^2\nabla\cdot(\rho_0 \vi) \rule{0mm}{5mm}\right. &-& \frac{{\bf B_0}}{4\pi}\cdot\nabla\cross\left(\frac{{\bf 
B_0}}{\rho_0}\cross\rho_0\vi\right)  \nn\\
&+& \left.\frac{1}{\rho_0}\frac{dP_{tot}}{dr}(\rho_0 \vi) \right] = 0. \label{eq:BCF4} \\
\eea

For completeness, we also study the stability of untruncated filaments, for which there is no external pressure.
When untruncated filaments extend radially to infinity, we simply do not include an internal boundary.  
We also consider the limit of an infinitely hot, zero density external medium.  In these cases, we assume that the external
medium is a non-conducting vacuum and use the boundary conditions prescribed by Nagasawa (1987).

\section{Matrix Representation of the Boundary Conditions}
\label{app:matrix}
The four boundary conditions given by equation \ref{eq:BCF1} to \ref{eq:BCF4} must now be inserted into the eigensystem given
by equation \ref{eq:EIG}.  Assuming that there are $N$ elements in our computational grid, with the interface between the molecular and
atomic gas between elements $I^\star$ and $I^\star+1$, we may write a matrix representation of our boundary conditions
in the form
\be
{\hat a}\Psi^\star+{\hat b}\Psi=0,
\label{eq:bcmat}
\ee
where
\be
\Psi^\star=\left[       \begin{array}{l}
                        \Psi_1          \\
                        \Psi_{I^\star}  \\
                        \Psi_{I^\star+1} \\
                        \Psi_N
                        \end{array}
        \right]
\ee
contains all of the boundary cells,
and $\Psi$ now contains the remaining $4(N-1)$ elements.  We likewise separate our eigensystem (equation \ref{eq:EIG}) into regular
and boundary grid cells:
\be
-\omega^2={\hat L}_1\Psi^\star+{\hat L}_2\Psi.
\label{eq:reg}
\ee
We may solve equation \ref{eq:bcmat} for the boundary values by elementary matrix algebra:
\be 
\Psi^\star=-{\hat a}^{-1} {\hat b} \Psi.
\ee
Substituting into equation \ref{eq:reg}, we eliminate 4 equations to obtain our eigensystem (of size $4(N-1)\times 4(N-1)$),
which takes into account all boundary conditions:
\be
-\omega^2\Psi=({\hat L}_2-{\hat L}_1 {\hat a}^{-1} {\hat b})\Psi.
\label{eq:EIGFINAL}
\ee
This is the final form of our eigensystem, which we solve in Section \ref{sec:disp}.

\label{lastpage}
\end{document}